\documentclass{emulateapj}
\usepackage{amssymb}
\usepackage{lscape}

\begin{document}
\newcommand{\lya}{Lyman~$\alpha$}
\newcommand{\lyb}{Lyman~$\beta$}
\newcommand{\degpoint}{\mbox{$^\circ\mskip-7.0mu.\,$}}
\newcommand{\minpoint}{\mbox{$'\mskip-4.7mu.\mskip0.8mu$}}
\newcommand{\secpoint}{\mbox{$''\mskip-7.6mu.\,$}}
\newcommand{\sqdeg}{\mbox{${\rm deg}^2$}}
\newcommand{\squig}{\sim\!\!}
\newcommand{\subsun}{\mbox{$_{\twelvesy\odot}$}}
\newcommand{\et}{{\it et al.}~}
\newcommand{\Rs}{{\cal R}}

\def\ltsima{$\; \buildrel < \over \sim \;$}
\def\simlt{\lower.5ex\hbox{\ltsima}}
\def\gtsima{$\; \buildrel > \over \sim \;$}
\def\simgt{\lower.5ex\hbox{\gtsima}}
\def\propsima{$\; \buildrel \propto \over \sim \;$}
\def\simprop{\lower.5ex\hbox{\propsima}}
\def\arcs{$''~$}
\def\arcm{$'~$}

\title{THE CONNECTION BETWEEN GALAXIES AND INTERGALACTIC ABSORPTION LINES AT REDSHIFT $2\simlt Z\simlt 3$\altaffilmark{1}}

\author{\sc Kurt L. Adelberger\altaffilmark{2}}
\affil{Carnegie Observatories, 813 Santa Barbara St., Pasadena, CA, 91101}

\author{\sc Alice E. Shapley\altaffilmark{3}}
\affil{University of California, Department of Astronomy, 601 Campbell Hall, Berkeley, CA 94720}

\author{\sc Charles C. Steidel}
\affil{Palomar Observatory, Caltech 105--24, Pasadena, CA 91125}

\author{\sc Max Pettini}
\affil{Institute of Astronomy, Madingley Road, Cambridge CB3 0HA, UK}
                                                                                
\author{\sc Dawn K. Erb \& Naveen A. Reddy}
\affil{Palomar Observatory, Caltech 105--24, Pasadena, CA 91125}

\altaffiltext{1}{Based, in part, on data obtained at the W.M. Keck
Observatory, which is operated as a scientific partnership between
the California Institute of Technology, the University of California,
and NASA, and was made possible by the generous financial support
of the W.M. Keck Foundation.}
\altaffiltext{2}{Carnegie Fellow}
\altaffiltext{3}{Miller Fellow}

\begin{abstract}
Absorption-line spectroscopy of 23 background QSOs and
numerous background galaxies
has let us measure the spatial distribution of metals and neutral
hydrogen around 1044 UV-selected galaxies at redshifts $1.8\simlt z\simlt 3.3$.
The typical galaxy is surrounded to radii $r\sim 40$ proper kpc
by gas that has a large velocity spread ($\Delta v> 260$ km s$^{-1}$)
and produces very strong absorption lines ($N_{\rm CIV}\gg 10^{14}$ cm$^{-2}$)
in the spectra of background objects.  
These absorption lines are almost as strong as those produced by a typical
galaxy's own interstellar gas.
Absorption with an average column density of $N_{\rm CIV}\simeq 10^{14}$ cm$^{-2}$ extends to
$\sim 80$ kpc, a radius large enough to imply that most strong
intergalactic CIV absorption is associated with star-forming galaxies
like those in our sample.
Our measurement of the galaxy/CIV spatial correlation function
shows that even the weakest detectable CIV systems are found
in the same regions as galaxies;
we find that the cross-correlation length increases 
with CIV column density
and is similar to the galaxy auto-correlation length 
($r_0\sim 4h^{-1}$ Mpc) for $N_{\rm CIV}\simgt 10^{12.5}$ cm$^{-2}$.
Distortions in the redshift-space galaxy-CIV correlation function on small
scales may imply that some of the CIV systems have large peculiar velocities.
Four of the five detected OVI absorption systems in our sample
lie within $400$ proper kpc of a known galaxy.  
Strong Lyman-$\alpha$ absorption is produced
by the intergalactic gas within $1h^{-1}$ comoving Mpc of most galaxies,
but for a significant minority
($\sim 1/3$) the absorption
is weak or absent.  This is not observed in SPH simulations
that omit the effects of ``feedback'' from galaxy formation.
We were unable to identify any statistically significant differences 
in age, dust reddening, environment, or kinematics
between galaxies with weak nearby HI absorption and the rest,
although galaxies with weak absorption may have higher star-formation rates.
Galaxies near intergalactic CIV systems appear to reside in relatively dense
environments and to
have distinctive spectral energy distributions that are
characterized by blue colors and young ages.
\end{abstract}

\submitted{Received 2005 February 07; Accepted 2005 May 09 }
\shorttitle{HIGH-REDSHIFT GALAXIES AND INTERGALACTIC ABSORPTION LINES}
\shortauthors{K.L. Adelberger et al.}

\section{INTRODUCTION}
\label{sec:intro}
A tremendous amount of gravitational energy is released
when a massive galaxy forms,
roughly $10^{62}$ erg from its billion supernova explosions
and another $10^{62}$ erg during the assembly of its central
$10^8$-$M_\odot$ black hole.  
Although $99$\% of the supernova energy is carried
away by neutrinos, and a similar fraction of the black hole
radiation escapes the galaxy, the remaining
energy is enough to unbind most of the galaxy's gas.
The fate of the galaxy depends on what happens to
the energy.
If it is
absorbed by sufficiently dense gas,
it will be converted into radiation
by two-body processes and will stream harmlessly away;
but otherwise it will set
in motion an enormous blast-wave
analogous to those produced by nuclear explosions on earth.
These blast-waves, tearing through galaxies,
stripping away material,
plowing outwards,
and coming finally to rest in distant intergalactic
space, are believed to have significantly altered the
evolution of the baryonic universe.

Indirect evidence for their existence
comes in two forms.  First, if the gas near galaxies had
not been heated by some source, disk galaxies would
be smaller than they are (Weil, Eke, \& Efstathiou 1998) and
the correlation between galaxy clusters' X-ray temperature
and luminosity would have the wrong shape (Kaiser 1991; 
Ponman, Cannon, \& Navarro 1999).  
Second, something quenches most
galaxies' star formation at some point in their history,
and blast-waves seem the best candidate.  Without them
small galaxies would be too numerous
(e.g., Cole et al. 1994), big galaxies
would be too blue (Springel, Di Matteo, \& Hernquist 2005)
and too large a fraction of baryons would have been turned into
stars (e.g., White \& Rees 1978, Springel \& Hernquist 2002).

Direct evidence remains elusive, however.  Although local star-forming galaxies are 
surrounded by outflowing gas (e.g., Heckman et al. 1990; Martin 2005), 
typically their observable outflows
extend to only a few kpc, far smaller than their virial
radii.  It is unclear how far these flows will advance before they stall
or how severely they will affect the evolution of the galaxies
and their surroundings.
The presence of metals in the intergalactic medium
might be taken as proof that the blast-waves spill out
of some galaxies' halos, but even that is controversial.
Intergalactic metals could also have been produced 
by the first
generation of Population III stars
or stripped from galaxies by other processes (e.g., Gnedin 1998).

Five years ago, our group began to search for direct
evidence for large-scale outflows around galaxies at high redshift.
High-redshift galaxies were attractive targets because they
have larger star-formation rates and (possibly) shallower potentials
than their nearby counterparts.  In addition, 
the higher density of the universe in the past
is a significant advantage, because it strengthens many of the
absorption lines that can be used to detect intergalactic gas.
We identified three observations that would qualify, in our view,
as direct evidence for large-scale outflows:  gas at large galactocentric
radii moving outwards at greater than the escape velocity; 
a strong association
of intergalactic metals and galaxies; and disturbances, near galaxies,
to the lattice-work of intergalactic HI that pervades
the high-redshift universe.

We searched for this evidence by systematically mapping the
relative spatial distributions of star-forming galaxies
and intergalactic gas.
Early results were presented in Adelberger et al. (2003).
This paper is an update.  After describing the
current data in \S~\ref{sec:data}, we review the status of
our search.
Section~\ref{sec:pairs} shows that the outflowing
material responsible for the blue-shifted ``interstellar'' absorption
lines in the galaxies' spectra often lies at a radius of 
$\sim 20$--$40 h^{-1}$ proper kpc.  Weaker absorption lines extend out to
radii approaching $200h^{-1}$ kpc.
Section~\ref{sec:r0gc} describes the galaxy-CIV cross-correlation
function and reinforces the idea that detectable intergalactic metals
tend to lie near actively star-forming galaxies.
Section~\ref{sec:gpe} shows that most galaxies are surrounded
by significant amounts of intergalactic HI but that galaxies with little
nearby HI are significantly more common than would
be expected in the absence of winds.
Section~\ref{sec:characteristics} discusses the characteristics
of galaxies that appear to be associated with unusually
weak Ly-$\alpha$ or unusually strong CIV absorption.
As we discuss in \S~\ref{sec:discussion}, these observations may be suggestive
of galactic winds, but they do not rule out other alternatives.
Our conclusions are summarized in \S~\ref{sec:summary}.
Throughout the paper we assume a cosmology with
$\Omega_M=0.3$, $\Omega_\Lambda=0.7$, $h=0.7$.

\section{DATA}
\label{sec:data}
The galaxies in our analysis were taken from the redshift surveys
of Steidel et al. (2003; 2004).  These surveys
targeted galaxies with magnitude ${\cal R}\simlt 25.5$ whose
$U_nG{\cal R}$ colors suggested that they had redshift
$2\simlt z\simlt 3$ (Adelberger et al. 2004; Steidel et al. 2003).  
We restricted our analysis to galaxies in the~13
survey fields that contained one or more background QSOs.
Although these fields include those previously analyzed by Adelberger et al.~(2003),
much of the present analysis relies on near-IR nebular redshifts (see below) that
are only available in other fields.  As a result, the overlap is small
between the samples analyzed here and in Adelberger et al.~(2003).

Our~13 fields are scattered around the
sky and have typical area $\Delta\Omega\sim 100$ arcmin$^2$ (Table~\ref{tab:fields}).
Redshifts for a total of~1044 objects in these fields
were measured from low-resolution ($\sim 10$\AA)
multislit spectra taken between 1995 and 2004
with LRIS (Oke et al. 1995; Steidel et al. 2004)
on the Keck I and II 10m telescopes.
The typical spectroscopic exposure time was 1.5 hours per multislit mask,
although a subset of the objects was observed for more than ten
times longer (Shapley et al. 2005, in preparation).
The redshifts obtained from these spectra are imprecise,
because they rely on features (Ly-$\alpha$ emission and
various far-UV interstellar absorption lines) that are
redshifted or blueshifted relative to the galaxy's stars.
Erb et al. (2005, in preparation) used NIRSPEC (McLean et al. 2000) on the Keck II telescope
to obtain more-precise redshifts from the nebular emission
lines ([OII]$\lambda 3727$, H$\alpha$, H$\beta$, [OIII]$\lambda\lambda 4959,5007$)
of 90 galaxies in this sample, many near the sightlines to background QSOs,
and we adopt their redshifts
wherever possible.  If nebular redshifts were unavailable,
we estimated the stellar redshifts with the relationships
defined in the following paragraph.  Precise redshifts were required
for much of the analysis, however, and in these cases we ignored
galaxies that had not been observed with NIRSPEC.

\begin{deluxetable}{lcrr
}\tablewidth{0pc}
\scriptsize
\tablecaption{Observed fields}
\tablehead{
        \colhead{Field} &
        \colhead{$\Delta\Omega$\tablenotemark{a}} &
        \colhead{$N_{\rm LRIS}$\tablenotemark{b}} &
        \colhead{$N_{\rm NIRSPEC}$\tablenotemark{c}}
}
\startdata
Q0000 & $3.4\times 4.5$ & 18 & 1 \\
Q0201 & $6.8\times 7.4$ & 25 & 3 \\
Q0256 & $7.3\times 6.6$ & 47 & 1 \\
Q0302 & $6.5\times 6.9$ & 44 & 1 \\
Q0933 & $8.1\times 8.0$ & 64 & 0 \\
Q1305 & $11.8\times 11.0$ & 80 & 1 \\
Q1422 & $7.2\times 14.2$ & 113 & 1 \\
Q1623 & $16.1\times 11.6$ & 213 & 31 \\
Q1700 & $11.5\times 11.0$ & 89 & 16 \\
Q2233 & $8.2\times 8.6$ & 46 & 1 \\
Q2343 & $22.5\times 8.5$ & 194 & 19 \\
Q2346 & $10.9\times 11.0$ & 50 & 7 \\
SSA22a & $7.8\times 8.4$ & 61 & 8 \\
\enddata
\tablenotetext{a}{Area imaged in $U_nG{\cal R}$ and observed spectroscopically,
arcmin$^2$}
\tablenotetext{b}{Number of galaxies with LRIS redshift $z>1$}
\tablenotetext{c}{Number of galaxies with NIRSPEC redshift $z>1$}
\label{tab:fields}
\end{deluxetable}

\begin{figure}
\plotone{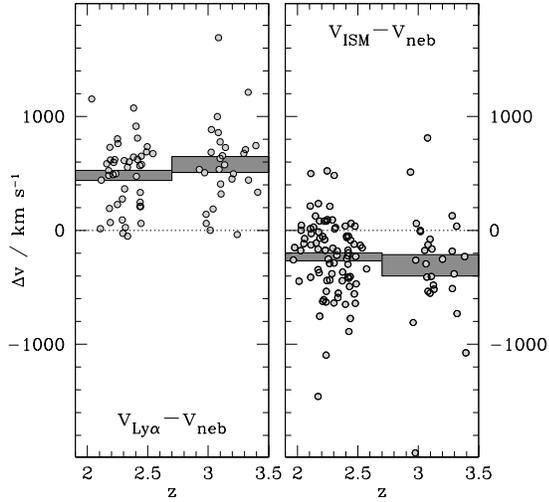}
\caption{ Velocity differences between the Lyman-$\alpha$ emission line,
interstellar absorption lines, and near-IR nebular lines
for galaxies in the NIRSPEC sample.  Each point represents one galaxy;
shaded regions show the mean velocity difference $\pm$ standard deviation of the mean
for two redshift bins.  The value of Spearman's rank correlation coefficient
and its significance are $r_s=0.16$, $P=0.19$ and $r_s=-0.197$, $P=0.03$
for the data in the left and right panels, respectively.  
The weak apparent correlations of velocity offset with redshift
are therefore significant at only the $1$--$2\sigma$ level.
\label{fig:check_dv}
}
\end{figure}

\begin{figure}
\plotone{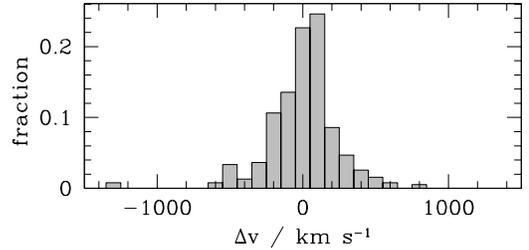}
\caption{Estimated histogram of velocity errors $\Delta v\equiv v_{\rm est} - v_{\rm true}$
for galaxies without near-IR nebular redshifts.
\label{fig:dvhistogram}
}
\end{figure}

We measured the relationship between redshifts from Lyman-$\alpha$ emission ($z_{{\rm Ly}\alpha}$),
interstellar absorption ($z_{\rm ISM}$), and near-IR nebular emission ($z_{\rm neb}$)
for the full 138-object
NIRSPEC sample of Erb et al. (2005, in preparation), and used the results
to help assign systemic redshifts to galaxies when NIRSPEC redshifts
were unavailable.  Figure~\ref{fig:check_dv} shows the observed velocity
offsets for the galaxies in the full NIRSPEC sample.  The least-squares
first-order fits to the data can be written
\begin{equation}
z_{\rm neb} = z_{{\rm Ly}\alpha} - 0.0033 - 0.0050(z_{{\rm Ly}\alpha}-2.7) 
\label{eq:em}
\end{equation}
for galaxies with no detected interstellar absorption lines,
\begin{equation}
z_{\rm neb} = z_{\rm ISM} + 0.0022 + 0.0015(z_{\rm ISM}-2.7)
\label{eq:abs}
\end{equation}
for galaxies with no detected Lyman-$\alpha$ emission, and
\begin{equation}
z_{\rm neb} = \bar z + 0.070\Delta z - 0.0017 - 0.0010(\bar z-2.7),
\label{eq:emabs}
\end{equation}
with $\bar z\equiv (z_{{\rm Ly}\alpha}+z_{\rm ISM})/2$, $\Delta z\equiv z_{{\rm Ly}\alpha}-z_{\rm ISM}$,
for galaxies with detectable Lyman-$\alpha$ emission and interstellar absorption.
The rms scatter around the three relationships is $\sigma_z = 0.0027$, $0.0033$, $0.0024$, respectively.
In each equation the final term accounts for the weak apparent trend of
smaller velocity offsets at lower redshifts.  Figure~\ref{fig:dvhistogram} shows
the expected error distributions for the galaxies
that lack near-IR nebular redshifts,
calculated by combining the
observed error distributions for equations~\ref{eq:em},~\ref{eq:abs}, and~\ref{eq:emabs}
after weighting by the number of galaxies in each class ($233$ with Lyman-$\alpha$ emission only,
$425$ with interstellar absorption only, $296$ with both).  
For comparison, repeated observations of galaxies in the NIRSPEC sample suggest that
these galaxies' redshifts have a random uncertainty of $60$ km s$^{-1}$.


The criteria used to select galaxies for spectroscopy were irregular.
We favored objects with $23<{\cal R}<24.5$ and objects near any known QSOs, and
strongly disfavored objects far from the field centers.  Near QSOs
we often observed objects whose colors satisfied any of the $U_nG{\cal R}$
selection criteria of Steidel et al. (2003) or Adelberger et al. (2004).
Far from QSOs our spectroscopic selection tended to be prejudiced towards
one or the other color-selection criterion, ``BX/BM'' or ``LBG,'' depending on the field.
This haphazard approach imposes a complicated selection bias on our data.
Rather than try to correct it with elaborate simulations, we
restrict our analysis to quantities that are not affected by
irregular angular sampling or the assumed redshift selection function.
Further details of the galaxy
observations can be found in Steidel et al. (2003; 2004)
and Erb et al. (2005, in preparation).

The QSO spectra we analyzed were obtained between 1996 and 2003
with either the HIRES echelle spectrograph (Vogt et al. 1994)
or the ESI echellette (Sheinis et al. 2000) on the Keck I and II
telescopes.  The HIRES spectra were reduced using the
standard procedures of Tom Barlow's Makee package.
The ESI spectra were reduced with the ``Dukee'' suite of custom 
IRAF scripts and C programs written by KLA and M. Hunt.
The shape of the continuum was removed from the QSO spectra
interactively by fitting a high-order ($\sim 30$) b-spline 
to regions of the spectra that appeared to be free from 
absorption.  CIV absorption systems in the QSO spectra
were identified interactively by scanning for absorption doublets
with the right wavelength spacing.  Column densities were estimated
by fitting each doublet's absorption profile with two Gaussians in optical depth.
This is equivalent to Voigt-profile fitting at the low (i.e., undamped) column-densities
of interest to us.
Occasionally the velocity profiles were unresolved in our spectra.
If this happened and the doublet ratio suggested that the lines were saturated,
we obtained a crude estimate of column density from the weaker line only.
Otherwise both lines were used.  See Adelberger et al. (2003)
for further details.  Our ability to detect CIV systems
depended on the quality of the QSO spectrum.  Table~\ref{tab:qsos}
lists the minimum detected CIV column density in each QSO's spectrum.
This should be roughly equal to the spectrum's CIV detection limit.

\begin{deluxetable*}{llrrccrc}
\tablewidth{0pc}
\scriptsize
\tablecaption{Observed QSOs}
\tablehead{
	\colhead{QSO} &
        \colhead{Field} &
	\colhead{$\alpha(2000)$} &
	\colhead{$\delta(2000)$} &
	\colhead{$z$} &
	\colhead{$G_{\rm AB}$\tablenotemark{a}} &
        \colhead{$n_{\rm CIV}$\tablenotemark{b}} &
        \colhead{${\rm log}(N_{\rm CIV}^{\rm min})$\tablenotemark{c}} 
}
\startdata
PKS0201+113\tablenotemark{d} & Q0201  & $02^h03^m46.7^s$ & $ 11^o34'45''$ & 3.610 & 20.1 &   0 & - \\
LBQS0256-0000    & Q0256  & $02^h59^m05.6^s$ & $ 00^o11'22''$ & 3.324 & 18.0 &  32 & 12.02 \\
LBQS0302-0019   & Q0302  & $03^h04^m49.9^s$ & $-00^o08'13''$ & 3.267 & 17.8 &  46 & 12.03 \\
FBQS J0933+2845 & Q0933  & $09^h33^m37.3^s$ & $ 28^o45'32''$ & 3.401 & 18.8 &  34 & 12.00 \\
Q1305kkc9       & Q1305  & $13^h07^m42.7^s$ & $ 29^o19'36''$ & 2.462 & 21.3 &   7 & 13.56 \\
Q1305kkc16      & Q1305  & $13^h08^m06.1^s$ & $ 29^o22'39''$ & 2.526 & 20.9 &   4 & 13.03 \\
Q1305kkc22      & Q1305  & $13^h08^m11.9^s$ & $ 29^o25'13''$ & 2.979 & 19.0 &  16 & 13.02 \\
Q1422+2309      & Q1422  & $14^h24^m38.1^s$ & $ 22^o56'01''$ & 3.515 & 16.5 &  74 & 11.81 \\
Q1422+2309b     & Q1422  & $14^h24^m40.6^s$ & $ 22^o55'43''$ & 3.084 & 23.4 &   3 & 13.45 \\
Q1623kp76       & Q1623  & $16^h25^m48.1^s$ & $ 26^o44'33''$ & 2.246 & 18.5 &  13 & 12.72 \\
Q1623kp77       & Q1623  & $16^h25^m48.8^s$ & $ 26^o46'59''$ & 2.529 & 16.5 &  41 & 12.03 \\
Q1623kp78       & Q1623  & $16^h25^m57.4^s$ & $ 26^o44'49''$ & 2.578 & 19.0 &  47 & 12.14 \\
Q1623kp79       & Q1623  & $16^h26^m06.2^s$ & $ 26^o50'33''$ & 2.180 & 19.2 &  12 & 12.99 \\
Q1623BX234      & Q1623  & $16^h25^m33.7^s$ & $ 26^o53'44''$ & 2.464 & 19.5 &   4 & 14.06 \\
Q1623BX603      & Q1623  & $16^h26^m02.4^s$ & $ 26^o43'39''$ & 2.529 & 20.5 &   8 & 13.32 \\
HS1700+6416     & Q1700  & $17^h01^m00.6^s$ & $ 64^o12'09''$ & 2.717 & 16.0 &  57 & 11.71 \\
Q2233+136       & Q2233  & $22^h36^m27.2^s$ & $ 13^o57'13''$ & 3.209 & 18.7 &  25 & 12.09 \\
Q2342+125       & Q2343  & $23^h45^m22.8^s$ & $ 12^o45'47''$ & 2.497 & 18.2 &   9 & 12.57 \\
Q2343+125       & Q2343  & $23^h46^m28.3^s$ & $ 12^o48'58''$ & 2.579 & 17.0 &  38 & 12.25 \\
Q2344+125       & Q2343  & $23^h46^m46.1^s$ & $ 12^o45'27''$ & 2.785 & 17.5 &  44 & 12.33 \\
Q2343BX415      & Q2343  & $23^h46^m25.4^s$ & $ 12^o47'44''$ & 2.578 & 20.3 &  19 & 13.22 \\
Q2345+000       & Q2346  & $23^h48^m25.4^s$ & $ 00^o20'41''$ & 2.652 & 19.7 &   3 & 13.31 \\
SSA22d13        & SSA22a & $22^h17^m22.3^s$ & $ 00^o16'41''$ & 3.343 & 21.6 &   6 & 13.22 \\
\enddata
\tablenotetext{a}{QSO magnitude; numbers are unreliable brighter than our saturation magnitude $G\sim 18$}
\tablenotetext{b}{Number of detected CIV systems}
\tablenotetext{c}{Log of the minimum column density (in cm$^{-2}$) among the detected CIV systems}
\tablenotetext{d}{CIV absorber catalog not constructed owing to gaps in our echelle spectrum}
\label{tab:qsos}
\end{deluxetable*}

\begin{figure}
\plotone{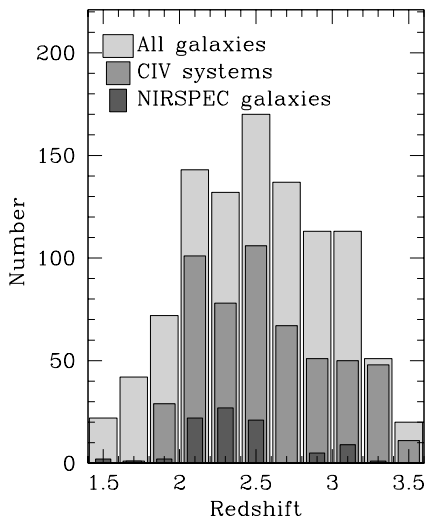}
\caption{ Redshift distributions for different objects in our sample.
\label{fig:nz}
}
\end{figure}

Figure~\ref{fig:nz} shows the redshift distributions of the different
objects in our sample.

\section{GALAXIES' GASEOUS ENVELOPES}
\label{sec:pairs}
Numerous authors have noticed that the interstellar absorption lines
in high-redshift galaxies' spectra tend to be blueshifted,
presumably because they are produced by outflowing gas
(e.g., Franx et al. 1997; Pettini et al. 1998, 2001;
Frye, Broadhurst, \& Ben\'itez 2002; Shapley et al. 2003; Erb et al. 2004).
Little is known, however, about how far the outflowing gas might
propagate.  This section shows that the absorbing material
often lies at radii approaching 80 proper kpc
and that weaker lines can be detected much farther away.
It does not clearly show, however, that the absorbing material 
detected at large radii is part of an outflow.  Other scenarios
are considered in \S\S~\ref{sec:r0gc} and~\ref{sec:summary}.

\subsection{Radii $r\simlt 80$ proper kpc}
\begin{figure}
\plotone{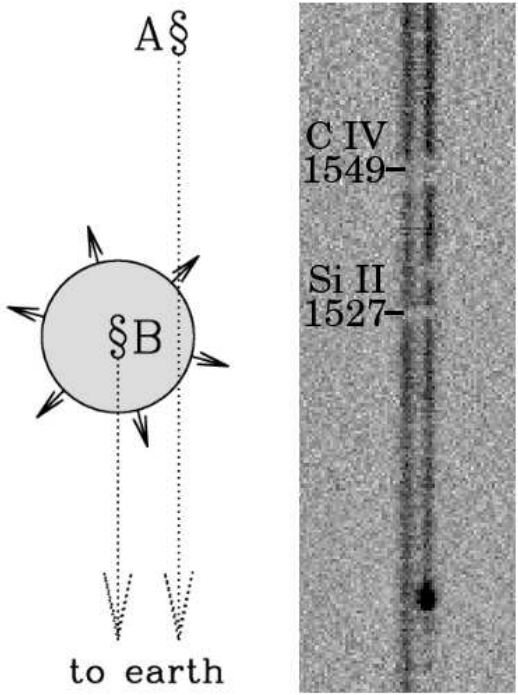}
\caption{
Left: Schematic view of galaxy/galaxy absorption.
A foreground galaxy, $B$, is surrounded by a gaseous
envelope that may be expanding.  Light from the
background galaxy, $A$, passes through this envelope
and is imprinted with absorption lines.
($A$ may have an expanding envelope of its own but
this is suppressed for simplicity.)
Right: Example LRIS spectra of a galaxy pair with
$2''$ separation and redshifts
$z=1.61$, $2.17$.  Spatial position varies along
the $x$ axis and wavelength increases towards
higher $y$.  The Lyman-$\alpha$ emission line of the
higher redshift object is visible at the bottom of
the rightmost spectrum.  The gas associated with the
lower redshift object, on the left, produces absorption
lines (labeled) in the spectra of both objects.  Its
radius must therefore exceed 17 kpc (i.e., $2''$
at $z=1.61$).
\label{fig:schematic}
}
\end{figure}

A $\sim 40$ kpc radius for galaxies' strongly absorbing material
is implied by the presence of foreground galaxies'
absorption lines in the spectra of nearby background galaxies.
The left panel of Figure~\ref{fig:schematic} 
illustrates the idea.  The right panel shows one example from our survey.
Two galaxies with redshifts $z=1.60$, $z=2.17$ and
angular separation $\theta=2''$
were placed on the same slit on one of our multislit masks.
Light from the background galaxy passed within
17 proper kpc of the foreground galaxy as it traveled towards earth,
and this was close enough for the foreground galaxy's
absorption lines to be imprinted in its spectrum.
The absorbing material must therefore have a radius of
at least 17 kpc.  

Similar absorption lines are observed in galaxy pairs with
impact parameters of up to 40 kpc, but the absorption 
weakens significantly at greater separations.
This can be seen by measuring the dependence on angular separation $\theta$ of the 
the mean absorption produced by foreground galaxies.
We calculated the strength of the mean absorption in two bins of angular separation.
For each bin, we identified all
the galaxy pairs in our sample (18 for $1''<\theta<5''$,
64 for $5''<\theta<10''$; $10''$ corresponds to 81 proper kpc
at $z=2.5$ for a cosmology with $\Omega_M=0.3$,
$\Omega_\Lambda=0.7$, $h=0.7$), collected the background spectra,
masked parts of the spectra that were contaminated by sky lines
or by any of the background galaxy's ten strongest spectral 
features\footnote{Ly$\beta$/OVI, Ly$\alpha$, SiII $\lambda 1260$,
OI $\lambda 1302$, CII $\lambda 1335$, SiIV $\lambda 1398$,
SiII $\lambda 1527$, CIV $\lambda 1549$, FeII $\lambda 1608$, 
AlII $\lambda 1671$},
shifted each spectrum into the rest-frame of its foreground galaxy,
spline-interpolated the rest-frame spectra onto common
abscissae, removed the shape of the continuum by forcing
each spectrum to have a constant 140\AA\ running-average flux,
and averaged the spectra together with inverse-variance weighting
after first rejecting the highest and lowest
10\% of the flux values at each output wavelength.
Finally we restored a realistic continuum shape by
multiplying the resulting spectrum by the mean running-average continuum shape
of the sample.

\begin{figure}
\plotone{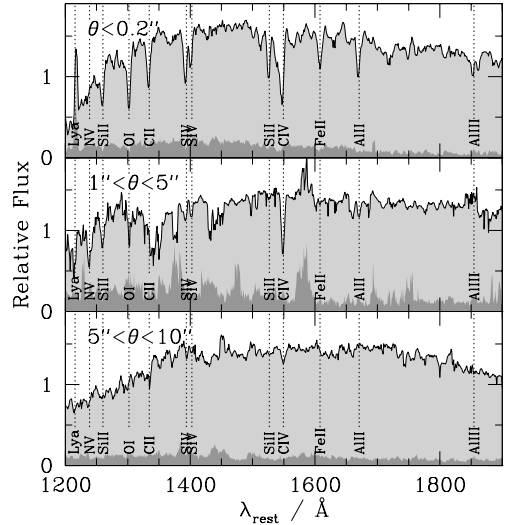}
\caption{
Mean gas absorption as a function of galactic impact parameter.
Top panel:  mean spectrum of the foreground galaxies in 
pairs with angular separation $1''<\theta<5''$
($8\simlt b\simlt 40$ proper kpc).  
Their absorption lines are sensitive to
the gas density at impact parameters smaller than their
half-light radii, i.e., $\theta\simlt 0.2''$.
The names of different absorption and emission lines are
indicated.  The darker shaded region at the bottom of the panel
is the bootstrap error spectrum, estimated by recalculating the mean spectrum
for random ensembles of foreground galaxy spectra.
Middle panel:  mean absorption observed in the spectrum
of the background galaxy at the redshift of the 
foreground galaxy.
This panel is noisier, owing to our masking of interstellar features
and (at bluer wavelengths) to Lyman-$\alpha$ forest absorption
in the spectra of the background galaxies.
Apparent features should be viewed skeptically.
Comparison to the error spectrum shows that some of
them are not significant
(e.g., absorption near 1375\AA, emission near 1580\AA).
Bottom panel: similar to the middle panel, but this
is the mean absorption produced in the spectra of
background galaxies with angular separation
$5''<\theta<10''$ ($40\simlt b\simlt 80$ proper kpc).
\label{fig:specsum}
}
\end{figure}

Figure~\ref{fig:specsum} shows the result.
The strength of the interstellar CIV absorption line appears roughly 
constant for $\theta\simlt 5''$ ($\sim 40$ proper kpc 
at $z=2.2$) but drops sharply afterwards.\footnote{Note
that part of the CIV feature in the top panel of Figure~\ref{fig:specsum}
is {\it stellar} P-Cygni absorption.}  
Chen, Lanzetta, \& Webb (2001) found similar results at lower
redshifts. 
It would be wrong to conclude from Figure~\ref{fig:specsum}, however,
that the observed CIV absorption at $5''<\theta<10''$ is
weak to the point of insignificance.  The estimated rest-frame
equivalent width of the absorption, $W_0({\rm doublet})\sim 0.7$\AA\
(i.e., $W_0(1548)\simgt 0.4$\AA, or $N_{\rm CIV}\simgt 10^{14}$ cm$^{-2}$),
is large by intergalactic standards.  According to Steidel (1990),
only $1.2\pm 0.4$ CIV systems with $W_0(1548)>0.4$\AA\
are found per unit redshift at $2\simlt z\simlt 3$.   Since our
color-selection criteria find 9 galaxies per arcmin$^2$ per unit z
at the same redshifts (e.g., Steidel et al. 2004), they would account
for the majority of all CIV-systems with $W_0(1548)>0.4$\AA\
if each galaxy produced absorption with $W_0(1548)>0.4$\AA\
out to $\theta=10''=0.167'$:  $9\times\pi (0.167')^2 \simeq 0.8$ CIV absorbers
per unit redshift.  In fact, as we will see below (Figure~\ref{fig:civhist}),
only about half of the galaxies within $10''$ ($\sim 80$ kpc) are
associated with CIV absorption so strong.  This implies that the galaxies
in our sample can account for roughly one-third of CIV lines 
with $W_0(1548)>0.4$\AA. 
Galaxies missed by our UV-color-selection techniques
could easily account for the remainder.

%\clearpage

The large equivalent width $W_0({\rm doublet})\simeq 2.7$\AA\
of the absorption at $\theta<5''$
has an interesting implication.  CIV can attain so high an equivalent
width only if peculiar velocities spread the absorption over a large
range of wavelengths.  The minimum velocity spread 
is $\Delta v = 260$ km s$^{-1}$, but
this assumes a contrived situation where each line in the doublet
is saturated and has a boxcar (ie, maximally compact) absorption profile.
In realistic situations the CIV absorption will have complicated substructure
(e.g., Pettini et al. 2002)
and a significantly larger velocity spread will be required to 
produce $W_0({\rm doublet})\simeq 2.7$\AA.
Our lower limit to the velocity spread ($\Delta v = 260$ km s$^{-1}$)
is smaller than the likely escape velocity at $40$ kpc
from a galaxy with $M\sim 10^{12} M_\odot$ (e.g., Adelberger et al. 2004),
but it nevertheless shows that significant chaotic
motions in the gas extend to large radii.

\begin{figure*}
\plotone{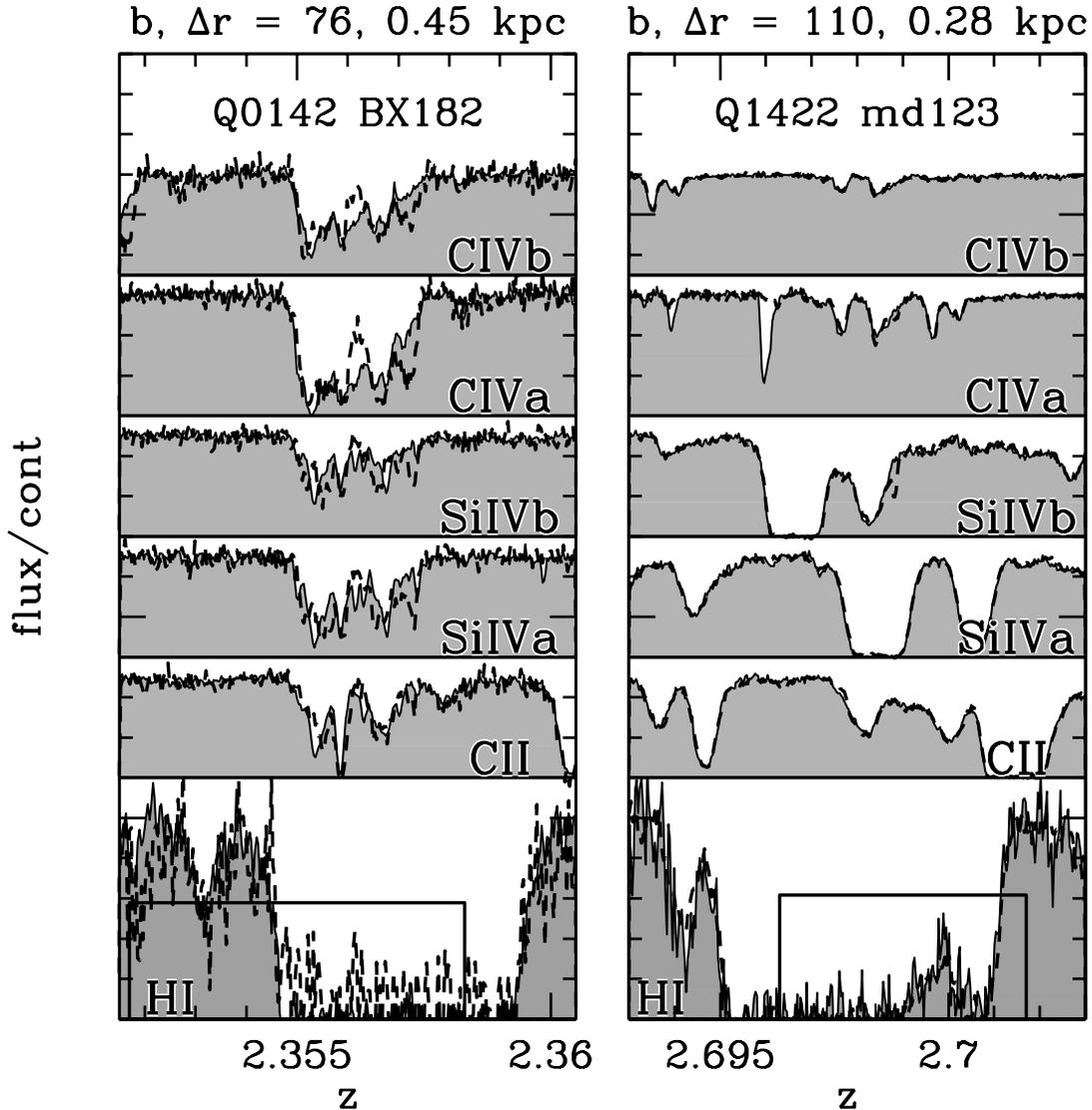}
\caption{
Substructure in metal-enriched gas near galaxies.
Each panel is for a single galaxy that lies
within $15''$ of a background QSO that is lensed into
two components with $\sim 1''$ separation.
The proper projected distance from galaxy to QSO sightline ($b$)
and proper projected separation of the two QSO components ($\Delta r$)
are indicated above the panels; $\Delta r$ is taken from Rauch et al. 2001a.
The $\pm1\sigma$ range of the galaxy's redshift is
indicated by the rectangle at the bottom of each figure.
The Lyman-$\alpha$ and metal-line absorption near the galaxy's redshift
is shown in the stacked spectral segments, which are drawn shaded
for one QSO component and dashed for the other.  
Significant differences
are apparent in the left-hand panel but not the right.
(In the right hand panel, the Lyman-$\alpha$ forest falls in the same 
part of the QSO spectrum as the SiIV and CII absorption lines,
making it difficult to determine if these features are present or absent.)
The recently obtained galaxy redshifts near Q0142-100 have been used
only for this plot, and this field has consequently been omitted from
Tables~\ref{tab:fields} and~\ref{tab:qsos}. 
\label{fig:smallscalestructure}
}
\end{figure*}

Further evidence comes from one of the two cases in our survey where
a foreground galaxy lies within $15''$ of the QSO sightline and
the QSO is lensed into two detectable components with $\sim 1''$ separation.
In this case, shown in the left panel of Figure~\ref{fig:smallscalestructure},
the galaxy's gas produces significantly different absorption profiles
in the spectra of the two QSO components.  The implied substructure
on half-kpc scales would be erased by thermal motions in a few dozen Myr
if it were not maintained somehow.  The gas
along the two sightlines might have supersonic relative motions
or might be gravitationally confined
in two separate mini-halos.
Substructure of this sort
is rare in the low-density intergalactic medium (Rauch et al. 2001b)
but common among CIV systems (Rauch, Sargent, \& Barlow 2001a).

\subsection{Radii $r\simgt 80$ proper kpc}
\begin{figure*}
\plotone{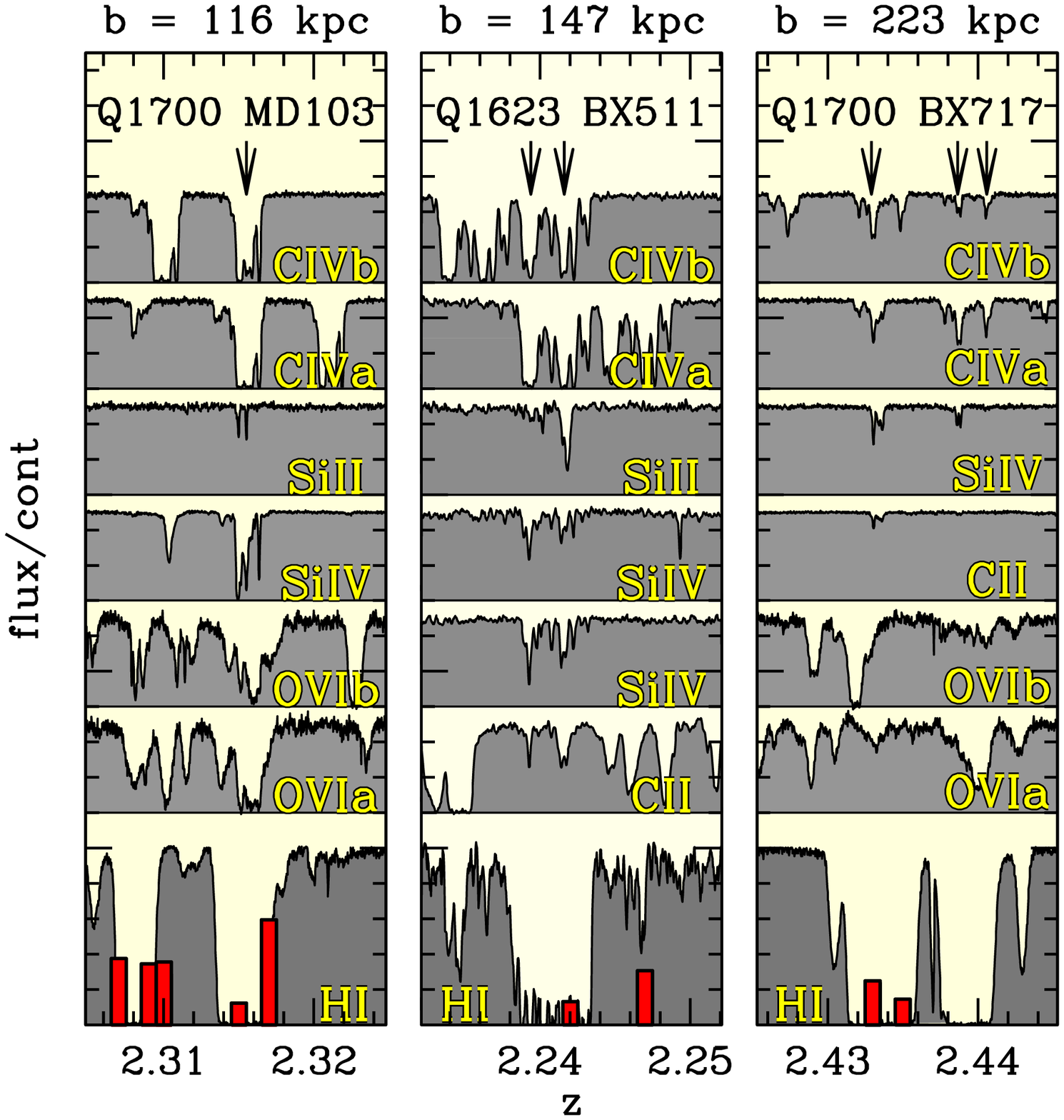}
\caption{
Absorption lines in QSO spectra at the redshifts of nearby galaxies.
Each of the three panels (left, middle, right) is for a single
galaxy, with proper impact parameter indicated above.
The galaxy's redshift, measured with NIRSPEC,
is marked by the short bar in the center of
the panel.  The width of the bar, $\pm 60$ km s$^{-1}$,
is equal to the redshift's $1\sigma$ uncertainty.
The redshifts of neighboring galaxies are indicated
with the other bars.  The neighboring galaxies were not observed with NIRSPEC
and so their redshifts are more uncertain, with $\pm 1\sigma$ limits
roughly 5 times broader than the bars.  
The bars' heights are proportional to the
galaxies' impact parameters.  The HI absorption near the galaxy's
redshift is shown at the bottom of the panel.  
Accompanying metal-line absorption, if any, is shown above.
Redshifts of detected metal absorption are marked with downward pointing
arrows near the top of the panel.
The velocity range shown in each panel, roughly $\pm 900$ km s$^{-1}$,
is small enough that absorption lines visible anywhere in a panel could
conceivably be produced by gas flowing out of the galaxy in the middle.
Some of the lines (e.g., OVI, CII) fall within the Lyman-$\alpha$ forest,
and observed absorption can be attributed to them with confidence
only if other metal-lines are present at the same redshift.
\label{fig:galmetals}
}
\end{figure*}

Although the metal absorption lines continue to weaken beyond $r\sim 80$ kpc,
they are not always absent. 
High signal-to-noise QSO spectra reveal that absorbing gas extends out to
$100$-$200$ proper kpc in at least some cases.
Figure~\ref{fig:galmetals} shows three.
Absorption from metal-enriched, multiphase gas is detected in each.
(See \S\S~2.5.13 and~2.5.15 of Simcoe et al. 2002 for a discussion of the physical conditions
in two of these absorption systems.)

\begin{figure}
\plotone{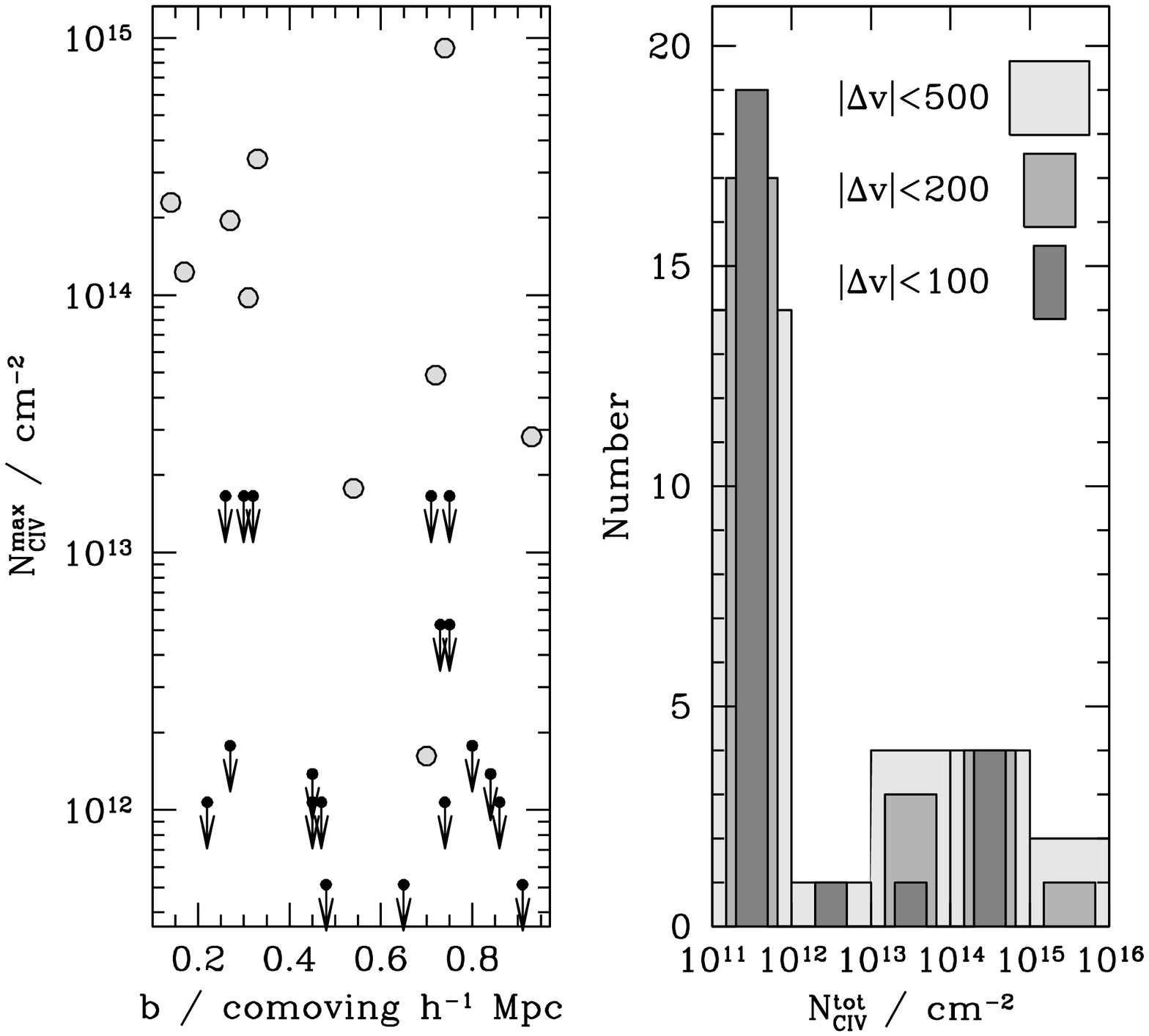}
\caption{
Left panel:  Column density of the strongest detected CIV system
within $\pm 200$ km s$^{-1}$ of the galaxy as a function of galaxy impact
parameter $b$.  Large points show detections.  Small points with
arrows show upper limits.  We assumed that the upper limit would
be equal to the column density of the weakest detected CIV system
in the QSO's spectrum.
Right panel:  Distribution of the total detected CIV column density
near galaxies that lie within $1h^{-1}$ comoving Mpc
of the QSO sightline.  The total column density
was calculated by summing the column densities of
all CIV systems within $100$, $200$, or $500$ km~s$^{-1}$
of the galaxy's redshift.  For simplicity, the non-detections
are all placed in the bin with $N_{\rm CIV}<10^{12}$ cm$^{-2}$.
This panel provides some indication of how the left panel
would change for velocity limits other than $\pm 200$ km s$^{-1}$.
The same galaxies
were used in both panels; all have
precise (i.e., near-IR nebular) redshifts.
\label{fig:civhist}
}
\end{figure}

These three cases are not typical, but nor are they
extraordinarily rare.  Figure~\ref{fig:civhist}
shows the total detected CIV column density
within 100--500 km s$^{-1}$ of the 29 galaxies in our
NIRSPEC sample that lie within $1.0h^{-1}$ projected comoving Mpc
of a QSO with detected CIV\footnote{A small fraction of our
QSOs were too faint to allow us to detect significant numbers
of CIV systems.}.
The CIV column densities of the three examples
in Figure~\ref{fig:galmetals} are the second, third, and seventh
highest in this 29 object sample.

One way to show that the detected metals are associated with
a nearby galaxy, and are not chance alignments,
is to measure the galaxy-CIV correlation strength $\bar\xi_{\rm gc}$
on small spatial scales.  As pointed out by Adelberger et al. (2003),
$\bar\xi_{\rm gc}$ cannot easily exceed $\bar\xi^{1/2}_{\rm gg}\bar\xi^{1/2}_{\rm cc}$,
the geometric mean of the galaxy-galaxy and CIV-CIV clustering strengths,
unless the existence of galaxies affects (or is affected by) the
presence of nearby CIV systems.  Since $\xi_{\rm gg}$ and $\xi_{cc}$
are roughly equal (Quashnock \& vanden Berk 1998; Adelberger et al. 2003), 
a value of $\bar\xi_{\rm gc}/\bar\xi_{\rm gg}$
significantly greater than unity on small scales
would imply a direct relationship between the galaxies and CIV systems.

We estimated $\bar\xi_{\rm gc}/\bar\xi_{\rm gg}$ with the following approach. 
For each CIV system, we counted the observed number $n_i$ of NIRSPEC galaxies 
with velocity separation $|\Delta v|<v_{\rm max}$ and 
transverse separation $R_{\rm min}<R<R_{\rm max}$.  
Call these galaxies the CIV system's neighbors.
(We took $v_{\rm max}=500$ km s$^{-1}$ and $R_{\rm max}\simlt 1h^{-1}$ comoving Mpc,
since these are roughly the maximum separations expected for a galaxy's ejecta.)
Given the galaxies' known angular distances
to the QSO, the expected number of neighbors
is (see, e.g., Adelberger 2005)
\begin{equation}
\bar n_i = \sum_j^{\rm galaxies}\frac{\int_{z_i-\Delta z}^{z_i+\Delta z} dz\, P(z) [1+\xi({\mathbf r_{ij}})]}{\int_{0}^{\infty} dz\, P(z) [1+\xi({\mathbf r_{ij}})]}
\label{eq:barn}
\end{equation}
where the sum runs over all NIRSPEC galaxies in the QSO's field,
$\Delta z$ is the redshift difference corresponding to $v_{\rm max}$,
$P(z)$ is the survey selection function normalized 
so that $\int_0^\infty P(z) = 1$, $\xi$ is the correlation function,
and ${\mathbf r_{ij}}$ is the distance between the CIV-system at redshift $z_i$
and a point at redshift $z$ with the galaxy's angular separation $\theta_{j}$.
The number of neighbors expected in the absence of clustering, $\bar n_i^0$,
can be calculated from equation~\ref{eq:barn} with $\xi=0$, allowing us to
estimate the galaxy-CIV clustering strength through $\bar\xi_{\rm gc} = -1+\sum_i n_i/\sum_i \bar n_i^0$.
This is the mean value of $\xi_{\rm gc}$ within a complicated volume consisting
of many parallel skewers of length $2v_{\rm max}$.
To estimate the mean value of the galaxy-galaxy correlation function within the same
volume, we inserted
$\xi=\xi_{\rm gg}=(r/4h^{-1}{\rm comoving\,\, Mpc})^{-1.6}$ (e.g., Adelberger et al. 2005)
into equation~\ref{eq:barn}, integrated numerically,
called the result $\bar n_i^{\rm gg}$, and used
the formula $\bar\xi_{\rm gg} = -1+\sum_i \bar n_i^{\rm gg}/\sum_i \bar n_i^0$.
The estimates of $\bar\xi_{\rm gc}$ and $\bar\xi_{\rm gg}$ 
do not depend on any assumptions about the redshift selection function of the CIV systems
or the angular selection function of the galaxies.  This is fortunate since
neither is well known.

\begin{figure}
\plotone{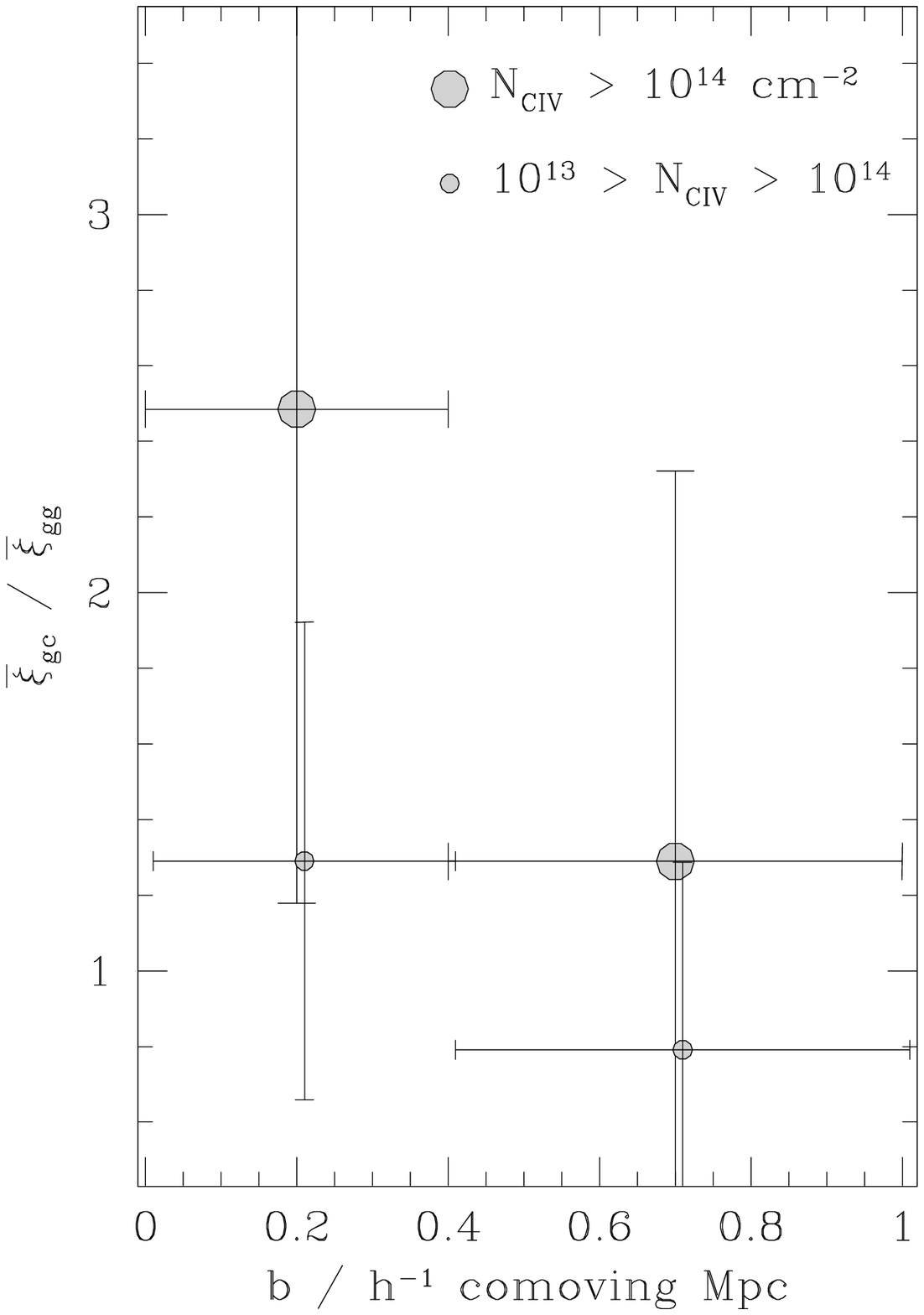}
\caption{
Estimated ratio of mean galaxy-CIV and galaxy-galaxy clustering strength
as a function of galaxy impact parameter $b$.  The correlation functions are averaged
over $\pm 500$ km s$^{-1}$ for each galaxy.
Points show the ratios for CIV systems in two column-density
ranges, as indicated.  
The $1\sigma$ error bars were calculated under the assumption
that the variance in the number of pairs $N$
is ${\rm Var}(N)\simeq 1.56N^{1.24}$, an approximation 
that Adelberger (2005) found to be accurate for galaxy pair
counts in the Lyman-break survey.
\label{fig:xiratio_vs_r}
}
\end{figure}

Figure~\ref{fig:xiratio_vs_r} shows $\bar\xi_{\rm gc}/\bar\xi_{\rm gg}$
for $R_{\rm min},R_{\rm max}$ = $0,0.4$ and $0.4,1.0 h^{-1}$ comoving Mpc.
Our data are consistent with no increase in $\bar\xi_{\rm gc}$
over $\bar\xi_{\rm gg}$ on small scales.  They are therefore
consistent with the idea that galaxies and CIV-systems are spatially
correlated only because they trace the same large-scale structure.
We cannot rule out a more direct association, however, both because
the error bars in Figure~\ref{fig:xiratio_vs_r} are large
and because the statistical test itself can detect only the strongest
direct associations of galaxies and CIV-systems.

\begin{figure}
\plotone{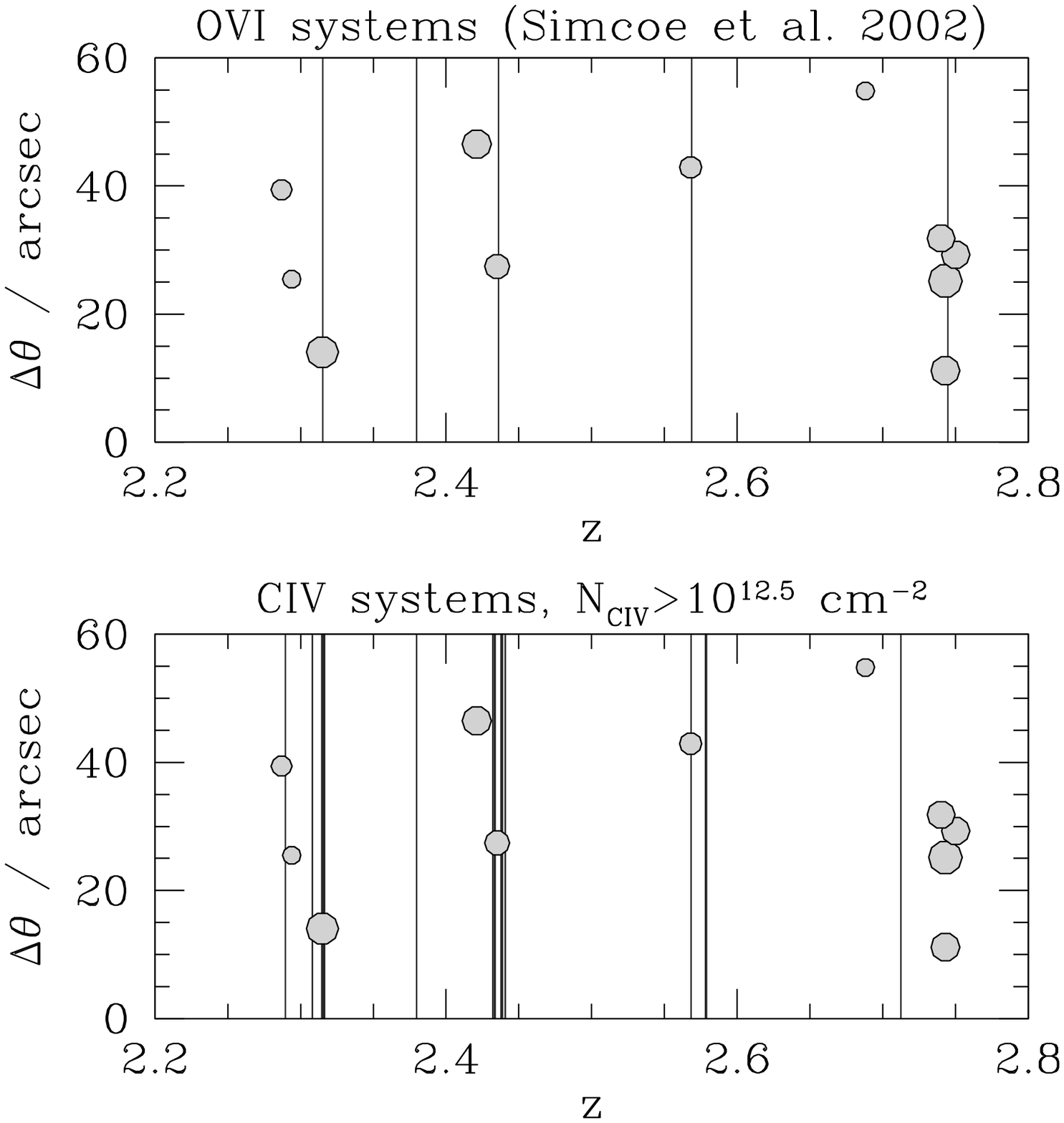}
\caption{
Redshifts of galaxies near the sightline
to QSO HS1700+6416 compared to the redshifts of
CIV and OVI absorption systems in the QSO's spectrum.
Circles mark galaxy redshifts and impact parameters.
The area of each circle is proportional to the galaxy's
apparent luminosity in the ${\cal R}$ band.
Vertical lines mark the redshifts of OVI and CIV absorption.
Thicker lines in the CIV panel mark systems with multiple components.
OVI absorption redshifts are taken from Simcoe, Sargent, \& Rauch (2002). 
The absorption line catalogs are reasonably complete down to
their limiting equivalent widths, but, 
owing to imperfect color-selection criteria and limited time for
follow-up spectroscopy, it is unlikely that the galaxy catalogs
contain more than half of the galaxies
brighter than ${\cal R}=25.5$ at these redshifts.
(We obtained redshifts for 15 of the 20
photometric candidates within $60''$ of the sightline to this QSO, so most of
the expected incompleteness is due to our color-selection criteria.
See, e.g., Adelberger et al. 2004.)
\label{fig:q1700metals}
}
\end{figure}

We conclude this subsection by emphasizing one of its corollaries:
many of
the strongest intergalactic absorption lines are produced
by material that lies 
close to star-forming galaxies.
Figure~\ref{fig:q1700metals} helps illustrate the point.
The figure's vertical lines mark the redshifts of strong CIV
and OVI absorption in the spectrum of HS1700+6416.\footnote{The
redshifts of OVI absorption were taken from Simcoe et al. (2002);
none of the other QSOs in our sample were analyzed by them, and
we have not attempted to find OVI systems ourselves.}
Circles
mark the positions of galaxies near the QSO sightline.
Small regions with a wealth of galaxies and intergalactic metals
tend to be surrounded by vacant expanses with neither.
The correspondence between star-forming galaxies and strong metal lines 
is not perfect, but one could do worse than assume that the
presence of one implies the presence of the other.

%\clearpage

\section{THE GALAXY-CIV CROSS-CORRELATION FUNCTION}
\label{sec:r0gc}

The previous section discussed the relationship between
galaxies and metal absorption lines for a small number of
specific cases.  This section takes a more systematic approach,
calculating the cross-correlation function 
between the positions of galaxies and
CIV absorption systems.  

\subsection{Large scales}

Our approach on large spatial scales is similar to that of Adelberger (2005):
we assume that the cross-correlation function has the form
$\xi(r) = (r/r_0)^{-1.6}$ (e.g., Adelberger et al. 2003); 
count the number $n_{\rm obs}(0,\ell)$
of galaxy-CIV pairs in our sample that have $\Delta\theta<300''$,
$|\Delta Z|<\ell$ for $\ell=20$, $40h^{-1}$ comoving 
Mpc\footnote{$\Delta Z$ is the
comoving separation in the redshift direction
between the galaxy and CIV system; i.e., it is the comoving distance
between the galaxy's redshift $z_g$ and the CIV system's $z_c$.};
calculate the expectation value
of $n_{\rm obs}(0,20)/n_{\rm obs}(0,40)$ as a function of $r_0$
with the equation
(Adelberger et al. 2005; Adelberger 2005)
\begin{equation}
\Biggl\langle\frac{n_{\rm obs}(0,\ell)}{n_{\rm obs}(0,2\ell)}\Biggr\rangle \simeq \frac{\sum_{i,j}\int_0^\ell dZ [1+\xi(R_{ij},Z)]}{\sum_{i,j}\int_0^{2\ell} dZ [1+\xi(R_{ij},Z)]},
\label{eq:k}
\end{equation}
where $\xi(R,Z)\equiv \xi[(R^2+Z^2)^{1/2}]$,
$R_{ij}\equiv (1+z_i)D_A(z_i) \theta_{ij}$, $D_A(z)$ is the angular diameter distance,
and the summation is over all galaxies ($i$) and CIV-systems ($j$);
and finally take as our best estimate of $r_0$ the value
that makes the right-hand side of equation~\ref{eq:k}
equal the observed ratio $n_{\rm obs}(0,20)/n_{\rm obs}(0,40)$.
As discussed in Adelberger et al. (2005) and Adelberger (2005),
the resulting estimate of $r_0$ is insensitive to angular selection 
effects and only weakly dependent on the assumed shape of the
selection function---two significant advantages given the irregular
selection criteria we employed (\S~\ref{sec:data}).

\begin{figure}
\plotone{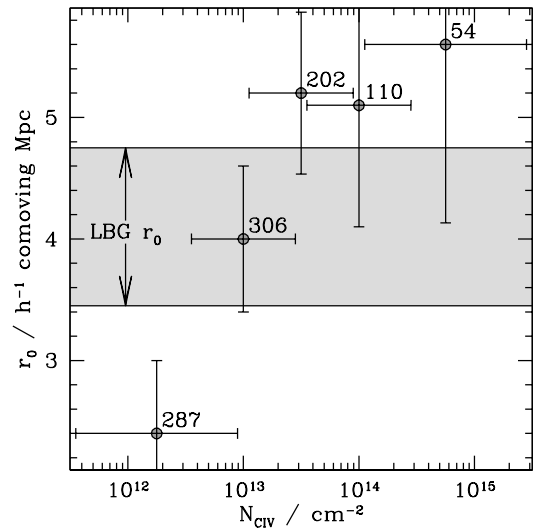}
\caption{
Galaxy-CIV cross-correlation length $r_0$ as a function
of CIV column density.  Circles show the measurements
of section~\ref{sec:r0gc}.  The number near each point
indicates the number of CIV systems with column densities
in the bin's range.  The shaded region in the
background shows the galaxy-galaxy correlation length $r_0$
appropriate to this sample (Adelberger et al. 2005).
\label{fig:r0gc_vs_nciv}
}
\end{figure}

Figure~\ref{fig:r0gc_vs_nciv} shows the estimated cross-correlation length
as a function of the CIV system's column density.  To estimate the uncertainty,
we broke our galaxy and CIV catalogs into 
many smaller sub-catalogs by rejecting a random fraction $p=0.5$--$0.8$
of the sources, observed how the dispersion in best-fit $r_0$
among the sub-catalogs depended on the number of objects in the catalogs,
and extrapolated to the full catalog sizes.   
This accounts for random uncertainties but not completely for
cosmic variance.

Also drawn in Figure~\ref{fig:r0gc_vs_nciv} is the $1\sigma$ confidence
interval for the galaxy-galaxy correlation length at similar
redshifts, taken from Adelberger et al. (2005).
The good agreement of the galaxy-galaxy and galaxy-CIV correlation
lengths for $N_{CIV}\simgt 10^{12.5}$ cm$^{-2}$
shows that bright star-forming galaxies and stronger CIV systems 
have similar spatial distributions, reinforcing the idea (\S~\ref{sec:pairs})
that they are often the same objects.  Adelberger et al.~(2003)
reached a similar conclusion.

\subsection{Small Scales}

\begin{figure*}
\plotone{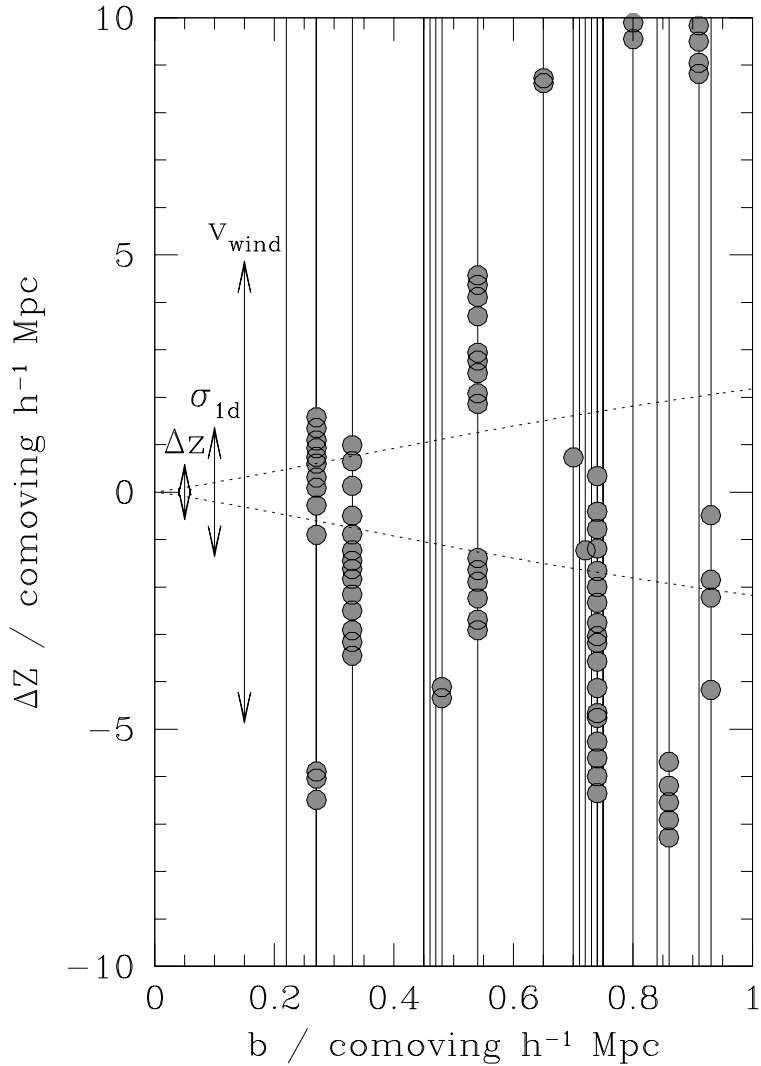}
\caption{
Spatial separations of galaxy-CIV pairs in the angular ($b$)
and redshift ($\Delta Z$) direction.  Vertical lines show the 
projected distance between each galaxy in the NIRSPEC sample
and its nearest QSO.  The QSO spectra allow us to detect
any CIV-absorbing gas along these lines.  Circles
show the actual locations of detected CIV absorption, with one
circle at the redshift of each resolved absorption component.
In the absence of peculiar velocities, the distribution of
CIV systems around galaxies would be spherically symmetric on average
and (as a result) half of the 
%Owing to the power-law shape of the galaxy-CIV correlation
%function, half of the
circles would be expected to lie inside the dotted triangular envelope
(see \S~\ref{sec:r0gc}).
%if there were no peculiar velocities.
In fact 65 of 81 lie outside.
The uncertainty in each galaxy-CIV pair's redshift separation (i.e., 
in each circle's vertical position) is too small
to account for this ($\pm1\sigma$ size indicated by two-headed arrow
labeled $\Delta z$).  
Peculiar velocities must therefore be substantial.
If the CIV systems were orbiting in the
galaxy's halo, their observed redshift separations would be displaced
relative to their true separations by a random amount comparable
to their galaxy's velocity dispersion $\sigma_{1d}$.  The approximate size
of $\pm\sigma_{1d}$ ($140$ km s$^{-1}$) is indicated by the labeled double-headed arrow.
If the CIV were infalling, the CIV systems' observed
(i.e., redshift space) positions would be artificially compressed towards
the $x$ axis by an amount comparable to $\sigma_{1d}$.
If it were flowing outwards, the observed positions would be displaced
away from the $x$ axis by an amount that depends on the wind speed.
The two-headed arrow labeled $v_{\rm wind}$
shows the approximate size of $\pm 500$ km s$^{-1}$.
\label{fig:civ_nirspec_xigc}
}
\end{figure*}

As mentioned above, an observed spatial association of galaxies and CIV systems
does not by itself imply that the detected metals are part of a galactic superwind.
Many other processes could place metals near galaxies.
One way to learn about the relationship between intergalactic CIV
and nearby galaxies is to look for anisotropies on small scales
in the redshift-space galaxy-CIV correlation function.  Since these anisotropies
are produced by peculiar velocities, they should reveal whether
the CIV tends to be falling towards, flowing away from, or orbiting around the nearest galaxy.
Figure~\ref{fig:civ_nirspec_xigc} shows the observed spatial separation
of every galaxy-CIV pair in our NIRSPEC sample that had comoving redshift separation
$|\Delta Z|<10h^{-1}$ Mpc and comoving angular separation $b<1h^{-1}$ Mpc.
Vertical lines mark the impact parameters of each galaxy-QSO pair.
Circles on these lines show the velocities (relative to the nearby galaxy) of
any detected CIV absorption in the QSO's spectrum.
Only galaxies with NIRSPEC redshifts are shown in the plot, since precise
redshifts are crucial to our arguments.  

The peculiar velocities of CIV systems relative to galaxies
can be estimated roughly from this plot as follows.
The expected number of galaxy-CIV pairs with impact parameter $b$
and radial separation $|\Delta Z|<Z_1$ is proportional to
$[1+\bar\xi(b,<Z_1)]Z_1$, where
\begin{equation}
\bar\xi(b,<Z_1) \equiv \frac{1}{Z_1} \int_0^{Z_1}dZ\,\xi(b,Z)
\label{eq:xibar}
\end{equation}
and $\xi(b,Z)$ is the galaxy-CIV correlation function.
On average, then, half of the plot's galaxies with impact parameter $b$
should have comoving redshift separation
$|\Delta Z|<Z_{1/2}(b)$ with
\begin{equation}
Z_{1/2}(b) = \frac{Z_{\rm max}}{2} \frac{1+\bar\xi(b,<Z_{\rm max})}{1+\bar\xi(b,<Z_{1/2})}
\label{eq:zhalf}
\end{equation}
where $|\Delta Z|<Z_{\rm max}$ is the range of redshift separations
shown on the plot.
If peculiar velocities were negligible,
$\xi(b,Z)$ would be equal to the real-space
galaxy-CIV correlation function $\xi^{\rm r}$, which must be isotropic
in an isotropic universe:
$\xi(b,Z) = \xi^{\rm r}[(b^2+Z^2)^{1/2}]$.
Since $\xi^{\rm r}(r)=(3.7 h^{-1} {\rm Mpc} / r)^{1.6}$
is a reasonably good approximation to the real-space correlation function
(see Figure~\ref{fig:xiratio_vs_r}), one can solve for
$Z_{1/2}(b)$ numerically by inserting this correlation function
into equations~\ref{eq:zhalf} and~\ref{eq:xibar}.

The dotted triangular envelope in Figure~\ref{fig:civ_nirspec_xigc}
shows $Z_{1/2}(b)$.  
Half of the CIV systems would be expected
to lie within the dotted envelope in the absence of peculiar velocities.
In fact 66 of 81 lie outside.  Peculiar velocities appear to be substantial.
The observations seem inconsistent with
an infall model, since infall tends to compress correlation functions in
the redshift direction; it would place more than half of the
CIV systems inside the envelope.  The random motions of orbiting clouds would
move some CIV systems outside the envelope through the
finger-of-god effect, but it is unclear if the expected size of the
displacements ($\pm 1.4h^{-1}$ comoving Mpc for $\sigma_{1d}=140$ km s$^{-1}$;
e.g., Adelberger et al. 2005) is large enough to account for our observations.
The small sample size prevents us from drawing firm conclusions, but
outflows moving at several hundred km s$^{-1}$ might become the favored
explanation
if the correlation function remains so anisotropic as the sample size grows.

\section{INTERGALACTIC HI NEAR GALAXIES}
\label{sec:gpe}
\begin{figure}
\plotone{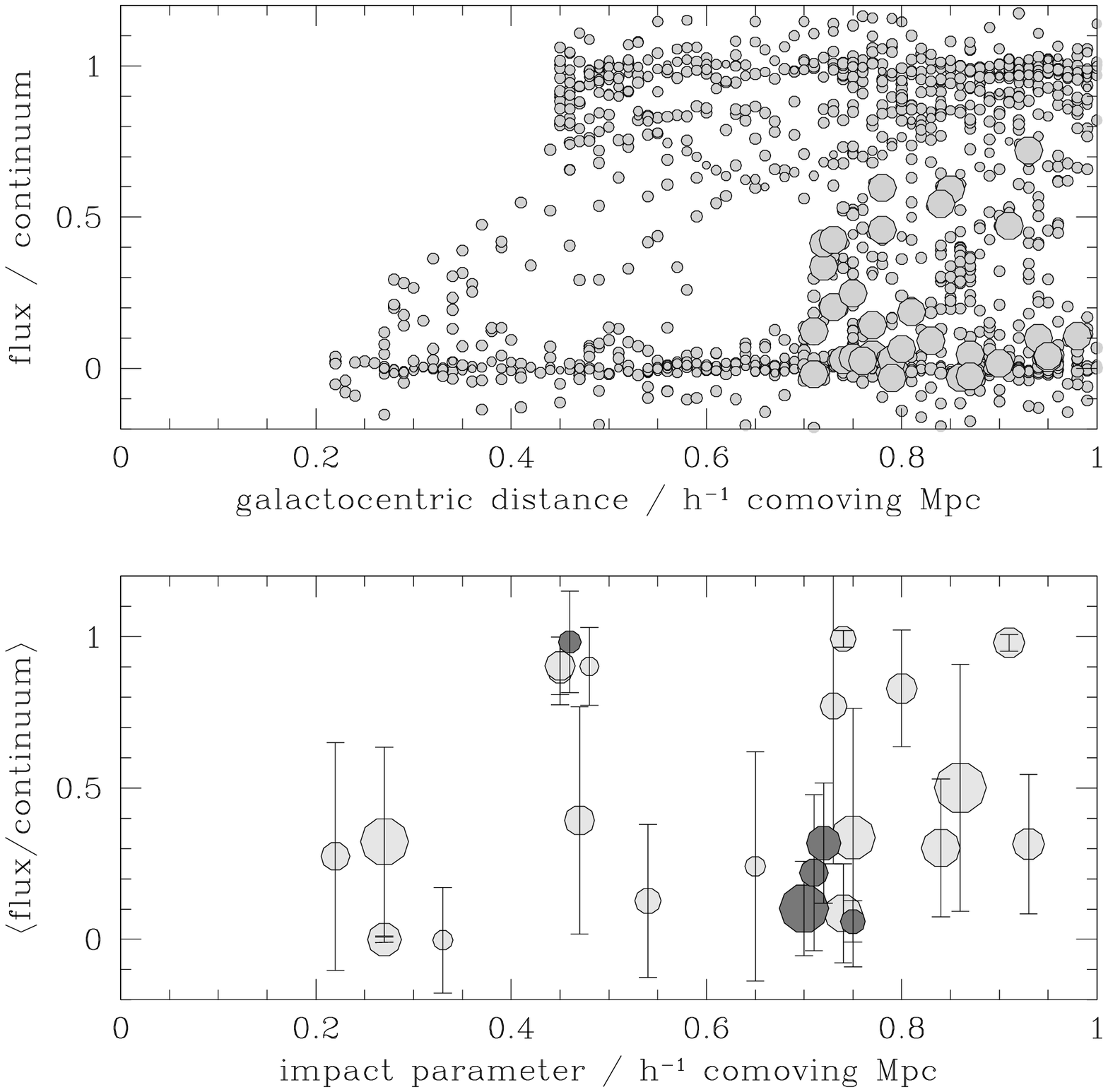}
\caption{
Top panel:  observed Lyman-$\alpha$ absorption for every QSO-spectrum pixel
whose apparent (i.e., redshift space)
distance to the nearest galaxy is $r<1h^{-1}$ 
comoving Mpc.  The circles are larger for lower-resolution
spectra; their area is proportional to the
wavelength size of the pixel and the smallest pixels have $\Delta\lambda\sim 0.025$\AA.
Bottom panel: the same data grouped by galaxy.  The mean transmissivity
for the pixels associated with each galaxy is shown on the $y$ axis.
Circles are placed on the $x$ axis according to the galaxy's impact parameter.
The area of each circle is proportional to the galaxy's apparent luminosity
in the ${\cal R}$ band.
Error bars show the rms spread in flux among the pixels
with $r<1h^{-1}$ Mpc.  Darker circles are galaxies from
Adelberger et al. (2003), re-analyzed after obtaining more-precise
galaxy redshifts from near-IR nebular lines.  We were able to obtain
confirming near-IR nebular redshifts for only one of the galaxies from that paper
with $r<0.5h^{-1}$ comoving Mpc; the other two had redshifts that
made near-IR spectroscopy impractical.
\label{fig:gpegals}
}
\end{figure}

After noticing that none of the three galaxies
closest to background QSOs in their sample
seemed to be associated with strong Lyman-$\alpha$ absorption lines,
Adelberger et al. (2003) speculated that superwinds
extending to $\sim 0.5h^{-1}$ Mpc might drive
most intergalactic hydrogen away from young galaxies.
Our larger sample puts this speculation to rest.
The top panel of Figure~\ref{fig:gpegals} shows
the measured transmissivity of every Lyman-$\alpha$ forest 
pixel\footnote{We excluded pixels within damped Lyman-$\alpha$ systems,
within $\Delta z=0.05$ of the QSO, or within parts of
the Lyman-$\alpha$ forest that could be contaminated by Lyman-$\beta$
absorption from gas at higher redshifts.}  in our QSO spectra
that lies within $1h^{-1}$ comoving Mpc of a galaxy
with precisely known (i.e., near-IR nebular) redshift.
The distances shown on the $x$-axis are imprecise, because
they were calculated from redshift differences under the 
assumption that peculiar velocities are negligible,
but nevertheless it is clear that
strong HI absorption is common near the galaxies.

The bottom panel of Figure~\ref{fig:gpegals} shows
the same data with the pixels grouped by galaxy.
Each point marks $\bar f_{1{\rm Mpc}}$, the mean Lyman-$\alpha$ transmissivity of the pixels
within $1h^{-1}$ comoving Mpc of a single galaxy, and the error
bar shows the r.m.s. spread among these pixel values.
Table~\ref{tab:gpegals} lists the impact parameters and redshifts
for the galaxies in this figure.  Also listed is $c$, the smoothed mean transmissivity
in the QSO's spectrum at the galaxy's redshift, calculated
by convolving the continuum-normalized QSO spectrum by a boxcar with width $100$\AA\ and
fitting a 2nd-order polynomial to the result.

\begin{deluxetable*}{llrrlrrrrrrrl
}\tablewidth{0pc}
\scriptsize
\tablecaption{Galaxies with impact parameter $b<1h^{-1}$ comoving Mpc\tablenotemark{a}}
\tablehead{
	\colhead{$b$\tablenotemark{b}} &
        \colhead{$z$\tablenotemark{c}} &
	\colhead{$\bar f_{1{\rm Mpc}}$\tablenotemark{d}} &
	\colhead{$c$\tablenotemark{e}} &
	\colhead{$N_{\rm CIV}$\tablenotemark{f}} &
	\colhead{$\Delta v_{\rm a}$\tablenotemark{g}} &
	\colhead{$\Delta v_{\rm e}$\tablenotemark{h}} &
	\colhead{$M_\ast$\tablenotemark{i}} &
	\colhead{SFR\tablenotemark{j}} &
	\colhead{$E$\tablenotemark{k}} &
	\colhead{Age\tablenotemark{l}} &
	\colhead{$\tau$\tablenotemark{m}} &
	\colhead{Galaxy name}
}
\startdata
0.14 & 2.181 &    - &    - & 14.36 &   19 &    x &    - &   - & -     &    - &    - & Q1623-BX215 \\
0.17 & 2.175 &    - &    - & 14.29 &    x &    x &  0.6 &  14 & 0.165 &  1.6 &  1.0 & Q2343-BX429 \\
0.22 & 2.182 & 0.27 & 0.84 &     x &    x &  500 &  1.5 &  12 & 0.060 & 12.8 &  100 & Q1623-BX432 \\
0.26 & 2.222 &    - &    - &     x & -130 &  419 &  2.4 &  15 & 0.110 & 16.1 &  100 & Q2343-BX417 \\
0.27 & 2.243 & 0.32 & 0.80 &     x & -435 &    x &  3.7 &  21 & 0.110 &  1.8 &  0.5 & Q2343-BX587 \\
0.27 & 2.315 & 0.00 & 0.79 & 14.92 &    x &    x & 10.5 &  19 & 0.160 &  4.0 &  1.0 & Q1700-MD103 \\
0.30 & 2.231 &    - &    - &     x & -167 &  529 &  1.2 &  31 & 0.150 &  4.0 &  100 & Q2343-BX390 \\
0.31 & 2.266 &    - &    - & 13.99 &    x &    x &    - &   - & -     &    - &    - & Q2346-BX120 \\
0.32 & 2.112 &    - &    - & 13.28 & -270 &  289 &  4.4 &  34 & 0.195 &  4.0 &  2.0 & Q2343-BX435 \\
0.33 & 2.242 & 0.00 & 0.75 & 15.11 &    x &    x &  1.2 &  10 & 0.180 &  2.5 &  1.0 & Q1623-BX511 \\
0.45 & 2.419 & 0.89 & 0.83 &     x &    x &    x &  2.3 &   5 & 0.010 &  6.4 &  2.0 & Q1623-BX449 \\
0.45 & 2.476 & 0.90 & 0.81 &     x & -130 &    x &  9.7 &  37 & 0.165 & 16.8 &   20 & Q1623-BX522 \\
0.46 & 2.931 & 0.98 & 0.67 &     - & -382 &    0 &    - &   - & -     &    - &    - & Q0201-md25 \\
0.47 & 2.148 & 0.39 & 0.83 &     x &  -29 &    x &  8.0 &  10 & 0.050 & 14.3 &  5.0 & Q1623-BX447 \\
0.48 & 1.968 &    - &    - &     - &    x &    x &  1.7 &   1 & 0.1   &  0.8 &  0.1 & Q2346-BX220 \\
0.48 & 2.294 & 0.90 & 0.78 & 12.81 &    x &    x &  2.9 &  14 & 0.170 & 18.0 &   50 & Q1700-MD109 \\
0.54 & 2.435 & 0.13 & 0.79 & 14.00 &    x &  236 &  0.7 &   4 & 0.005 &  2.9 &  1.0 & Q1700-BX717 \\
0.65 & 2.189 & 0.24 & 0.77 &     x &    x &    x & 13.7 &   9 & 0.125 & 27.5 &   10 & Q1700-BX691 \\
0.70 & 3.039 & 0.10 & 0.71 & 12.21 &    x &    x &    - &   - & -     &    - &    - & Q2233-MD39 \\
0.71 & 3.018 & 0.22 & 0.59 &     x &    x &    x &    - &   - & -     &    - &    - & SSA22a-C31 \\
0.72 & 3.099 & 0.32 & 0.60 & 13.69 & -278 &  381 &    - &   - & -     &    - &    - & SSA22a-C35 \\
0.73 & 2.059 & 0.77 & 0.61 &     x &   98 &    x &  9.3 &  16 & 0.135 & 27.5 &   20 & Q1623-BX452 \\
0.74 & 2.054 & 0.09 & 0.78 & 15.17 &  -30 &  511 &  2.0 &   2 & 0     &  2.5 &  0.5 & Q1623-BX428 \\
0.74 & 2.407 & 0.99 & 0.83 & 14.12 &  -70 &    x &  0.1 &  86 & 0.270 &  0.1 &  0.1 & Q1623-BX455 \\
0.75 & 2.016 & 0.29 & 0.56 &     x & -199 &    x &  0.6 &   4 & 0.070 &  0.5 &  0.1 & Q1623-BX429 \\
0.75 & 3.066 & 0.06 & 0.59 &     x &    x &  826 &    - &   - & -     &    - &    - & SSA22a-C39 \\
0.80 & 2.340 & 0.83 & 0.73 &     x & -323 &    x &  1.2 &  19 & 0.110 &  2.9 &  2.0 & Q2343-BX537 \\
0.84 & 2.424 & 0.30 & 0.81 &     x & -166 &    x &  9.2 &  51 & 0.145 & 18.0 &  100 & Q1623-BX516 \\
0.86 & 2.409 & 0.50 & 0.83 & 13.90 & -211 &  598 &  0.6 &  92 & 0.180 &  0.2 &  0.1 & Q1623-BX376 \\
0.91 & 2.421 & 0.98 & 0.79 &     x &    x &    x &  3.6 &  22 & 0.155 &  2.9 &  1.0 & Q1700-BX759 \\
0.93 & 2.174 & 0.32 & 0.81 & 14.26 & -151 &  331 &  2.6 &   9 & 0.010 & 27.5 &  100 & Q2343-BX660 \\
\enddata
\tablenotetext{a}{Unobserved quantities are marked ``-''; quantities we could not measure owing
to noise in the data are marked ``x.''}
\tablenotetext{b}{Comoving impact parameter, $h^{-1}$ Mpc}
\tablenotetext{c}{Redshift}
\tablenotetext{d}{Mean transmissivity of pixels within $1h^{-1}$ comoving Mpc}
\tablenotetext{e}{Mean transmissivity of random pixels with similar redshifts in the QSO's spectrum}
\tablenotetext{f}{Total detected CIV column density within 200 km s$^{-1}$ of the galaxy redshift, cm$^{-2}$}
\tablenotetext{g}{Velocity difference between interstellar absorption lines and nebular emission lines, km s$^{-1}$.  Typical uncertainty $\sim 200$ km s$^{-1}$.}
\tablenotetext{h}{Velocity difference between Lyman-$\alpha$ and nebular emission lines, km s$^{-1}$.  Typical uncertainty $\sim 100$ km s$^{-1}$.}
\tablenotetext{i}{Stellar mass, $10^{10} M_\odot$.  Typical uncertainty $\sim 0.15$ dex}
\tablenotetext{j}{Star formation rate, $M_\odot$ yr$^{-1}$.  Typical uncertainty $\sim 0.2$ dex}
\tablenotetext{k}{Dust reddening $E(B-V)$.  Typical uncertainty $\sim 0.03$}
\tablenotetext{l}{Age, $10^8$ yr.  Typical uncertainty $\sim 0.3$ dex}
\tablenotetext{m}{Time constant of exponentially declining star-formation history, $10^8$ yr.  Typical uncertainty $\sim 1$ dex}
\label{tab:gpegals}
\end{deluxetable*}

\begin{figure}
\plotone{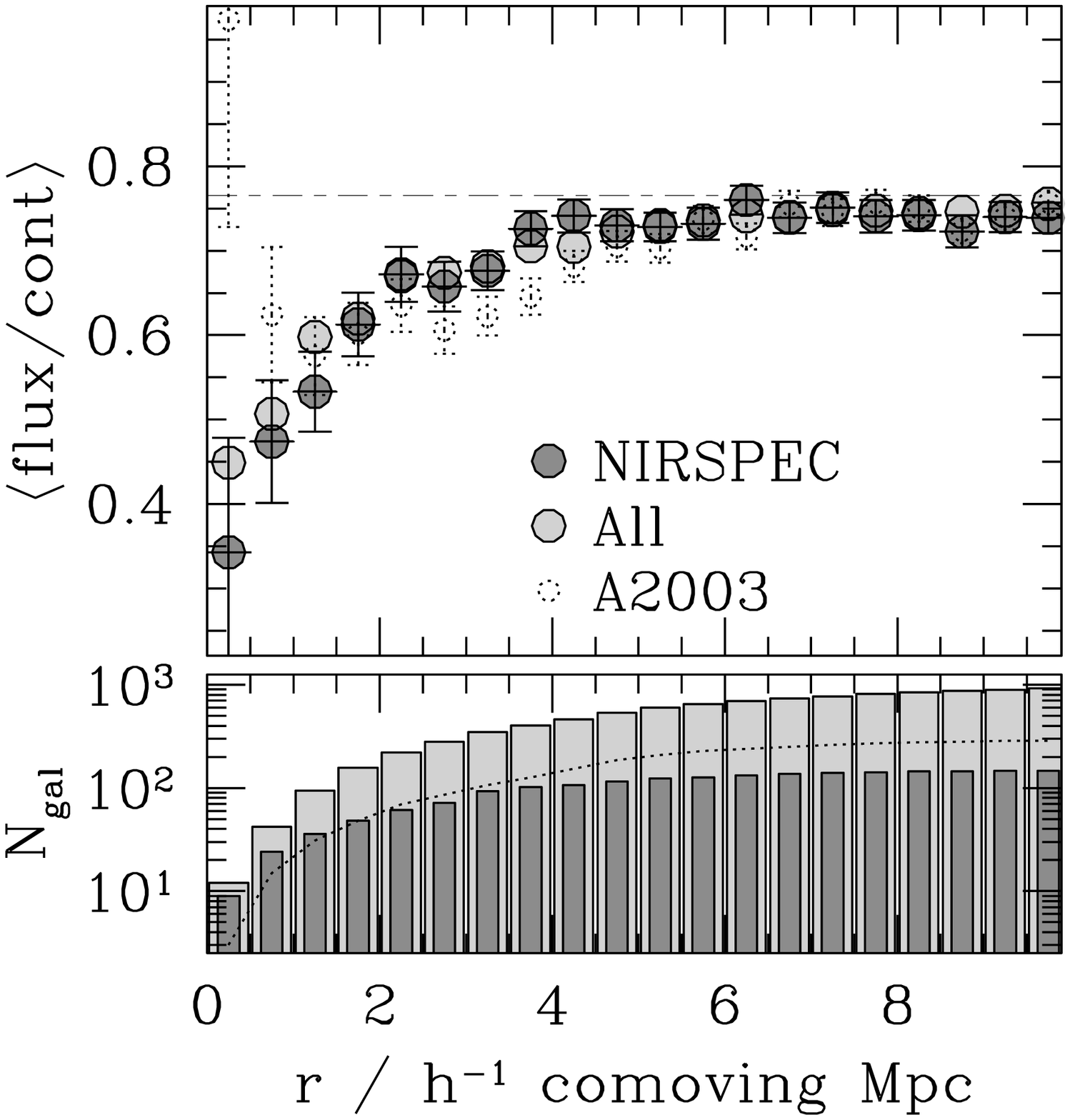}
\caption{
Top panel:
Mean transmissivity as a function of galactocentric distance.  Points
at $0<r<0.5h^{-1}$ comoving Mpc show the mean transmissivity of all spectral
pixels whose redshift-space position lay within $0.5h^{-1}$ comoving Mpc of a galaxy.
To scale to a common global transmissivity $\langle f\rangle = 0.765$, 
the transmissivities of each pixel were multiplied by $\langle f\rangle/c(z)$
before averaging,
with $c(z)$ the redshift-dependent smoothed mean transmissivity in each
QSO's spectrum (\S~\ref{sec:gpe}).
The error bar shows the $1\sigma$ uncertainty, estimated by
dividing the observed galaxy-to-galaxy scatter in transmissivity by $N_{\rm gal}^{1/2}$.
The points and error bars at larger separations are defined similarly.
Darker circles show the result if we restrict ourselves to galaxies
with near-IR nebular redshifts.  Lighter circles are for the full sample.
For comparison, dotted circles show the result of Adelberger et al. (2003)
scaled from $\langle f\rangle = 0.67$ to $\langle f\rangle = 0.765$.
Bottom panel:  the number of galaxies at each separation in each sample.
\label{fig:fbar_vs_r}
}
\end{figure}

Figure~\ref{fig:fbar_vs_r} shows that the mean transmissivity declines
monotonically as one approaches a galaxy.  
This appears inconsistent with the earlier result of Adelberger et al. (2003; dashed circles).
The present analysis differs from theirs in three main ways:  
our redshifts are more precise, our sample is larger, 
and our galaxies' typical redshift is lower.  
Each difference could contribute to the change in this plot.  
The good agreement of the
results for the NIRSPEC-only and full samples in Figure~\ref{fig:fbar_vs_r}
suggests that
the redshifts' precision might not be primarily responsible for the change.
This suspicion is reinforced by our NIRSPEC observations of galaxies
in the sample of Adelberger et al. (2003).  Although two of their
three galaxies with $r<0.5h^{-1}$ comoving Mpc had redshifts that
placed the nebular lines outside of atmospheric windows, the near-IR nebular
redshift we obtained for the third confirmed the reported lack of nearby Lyman-$\alpha$ absorption.\footnote{though it altered the galaxy's redshift by more than anticipated by Adelberger et al. (2003).}
Another possibility is that
genuine evolution from
$z\sim 3$ to $z\sim 2$ is partly responsible for the increased absorption
near galaxies in this sample.
The thought is not absurd:
at lower redshifts, galaxies have marginally slower outflows
(see Figure~\ref{fig:check_dv}) and a larger fraction of intergalactic
HI absorption is produced by gas in deep potential wells that is more
difficult to disrupt.   These changes are not enormous, however,
and it seems unlikely to us that they would have a dominant effect.
The most obvious difference 
from Adelberger et al. (2003) is the sample size,
especially at small separations.  The following simple argument
suggests that a statistical fluctuation could plausibly be responsible
for lack of absorption at $r<0.5h^{-1}$ comoving Mpc reported by Adelberger et al. (2003).
Suppose that half of the galaxies with $r<0.5h^{-1}$ comoving Mpc
are heavily obscured ($f=0$) and half are unobscured ($f=1$).
This is a rough approximation to the data in the bottom panel of Figure~\ref{fig:gpegals}.
Then the mean and standard deviation of the average transmissivity
for a three-object sample would be $\bar f_3=0.5\pm 0.29$.  
Although an observation of $\bar f_3=1$ superficially seems inconsistent
with this scenario, differing by almost $2\sigma$ from the expectation $\bar f_3=0.5$, 
in fact it should occur
one time out of eight.  The real world is only vaguely similar to our
simple model, but the point is that large fluctuations occur often in samples
drawn from a bimodal underlying distribution.   The results in this paper
and in Adelberger et al. (2003) are clearly not the same, but they
are less inconsistent than one might naively suppose.

\begin{figure}
\plotone{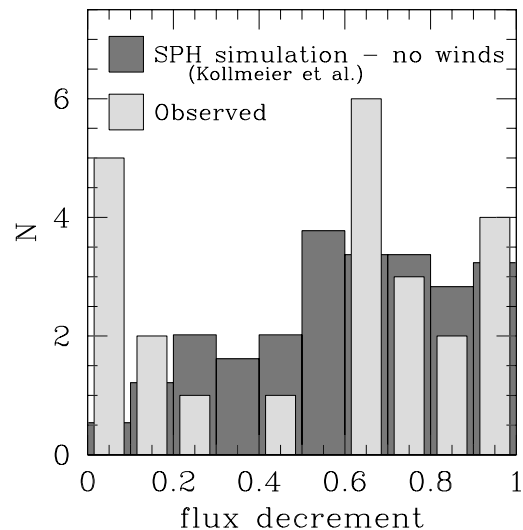}
\caption{
Comparison of observed and expected Ly-$\alpha$ absorption
near star-forming galaxies.  The abscissa shows the
flux decrement $1-\bar f_{1{\rm Mpc}}$, where
$\bar f_{1{\rm Mpc}}$, the mean transmissivity among all pixels within
$1h^{-1}$ comoving Mpc of the galaxy, is the quantity shown in the bottom
panel of Figure~\ref{fig:gpegals}.  In windless SPH simulations,
almost every sightline near a galaxy has strong Lyman-$\alpha$ absorption.
This is not observed.
\label{fig:compare_juna}
}
\end{figure}

In any case, 
although strong HI absorption is the norm at small separations,
we continue to find some galaxies with little absorption.
The existence of these galaxies is surprising:
one would expect all galaxies to reside in dense regions
with significant hydrogen and an elevated
neutral fraction from the $\rho^2$ dependence of the
recombination rate (e.g., Croft et al. 2002; Kollmeier et al. 2003; 
Maselli et al. 2004; Desjacques et al. 2004).
Figure~\ref{fig:compare_juna} illustrates this
point by comparing the observed
distribution of HI absorption among the galaxies
in Figure~\ref{fig:gpegals} to a preliminary prediction
from the windless smooth-particle hydrodynamical (SPH) simulation
of Kollmeier et al. (2005, private communication; this simulation
is similar to that of Kollmeier et al. 2003 but is matched
to the mean redshift of this sample).
If the real world resembled the simulations, 
a 24 galaxy sample would have $n_{\rm obs}=7$ or more members
with decrement $D<0.2$
only two times in a thousand
(i.e., $\sum_{n_{\rm obs}=7}^{\infty} \exp(-n_{\rm exp}) n_{\rm exp}^{n_{\rm obs}} / n_{\rm obs}!\simeq 0.002$, 
where $n_{\rm exp}\simeq 1.75$ is the predicted
mean number of galaxies with $D<0.2$ in a 24 object sample).

%\clearpage

Although a number of explanations are possible,
the data in Figure~\ref{fig:compare_juna} seem qualitatively
consistent with the idea that the galaxies are surrounded by
anisotropically expanding winds.
Leaking out along the paths of
least resistance, these winds would turn moderate absorption into weak
absorption while leaving the strongest absorption systems in place.

\begin{figure*}
\plotone{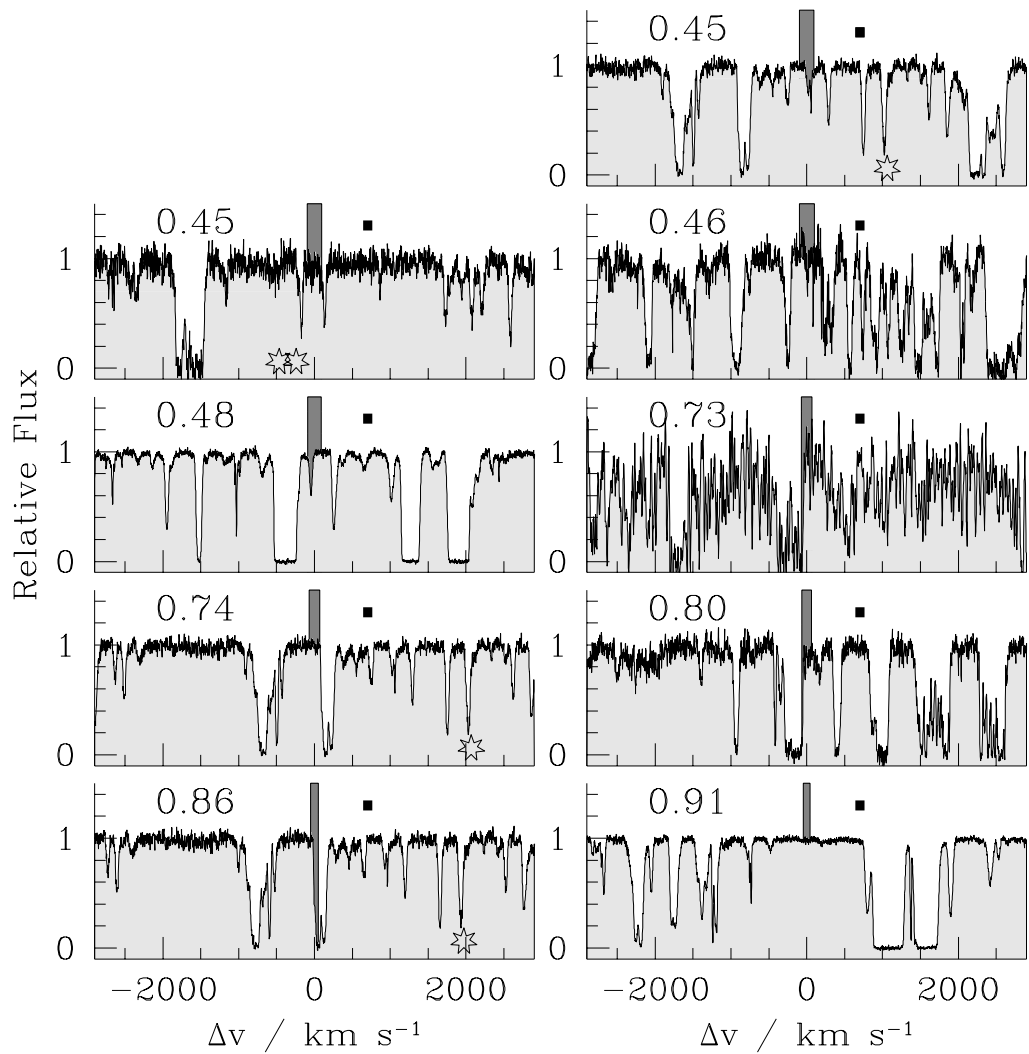}
\caption{
HI absorption near the redshifts of the 9 sightline
galaxies with the weakest absorption in Figure~\ref{fig:gpegals}.
The lighter shaded region shows the HI absorption
in the QSO's spectrum.  The darker shaded box in the background
is centered on the galaxy's redshift.  Its width
marks the range of pixels that lie within $1h^{-1}$ comoving Mpc of
each galaxy.  The number to the upper left of each panel gives
each galaxy's impact parameter in $h^{-1}$ comoving Mpc.
The redshifts of neighboring QSOs with $G_{AB}<20.5$ in each field
are marked with stars; the QSOs whose spectra are displayed
lie at redshifts too high to fall on this plot.
The neighboring QSO in the upper right panel and two lowest left panels
has $G_{AB}\sim 20.5$ and lies $\sim 9'$ (4 proper Mpc) from the sightline;
the two neighboring QSOs in the upper left panel have $G_{AB}=18.5$, $19.5$
and lie $\sim 100''$, $600''$ from the sightline.
The black square in each panel has a width of $120$ km s$^{-1}$,
twice the estimated $1\sigma$ uncertainty in each galaxy's redshift.
The largest positive and negative changes in redshift estimate
we have encountered
when re-observing a galaxy with NIRSPEC, 
$124$ km s$^{-1}$ and $-95$ km s$^{-1}$,
would displace a galaxy by about this distance.
\label{fig:check_gpegals}
}
\end{figure*}

Ionizing radiation from any neighboring QSOs might also preferentially
eliminate weaker absorption systems.  As Figure~\ref{fig:check_gpegals} shows,
many of the galaxies with $\bar f_{\rm 1{\rm Mpc}}$ have bright QSOs
in their vicinity.  Although these QSOs have similar or larger redshifts
than the galaxies, as is required for their (presumably) shortlived radiation to have an
observable effect on the galaxies' surroundings (e.g., Figure~12 of Adelberger~2004),
the prevalence of neighboring QSOs is not significantly
different for galaxies with $\bar f_{1{\rm Mpc}}\geq 0.5$
and $\bar f_{1{\rm Mpc}}<0.5$.
Four galaxies of~9 with $\bar f_{1{\rm Mpc}}\geq 0.5$
have a QSO within $\Delta v=3000$ km s$^{-1}$, compared
to 2 of 15 with $\bar f_{1{\rm Mpc}}<0.5$.  The one-sided $P$-value 
from Fisher's (1934) exact test is 0.11,
and would increase (i.e., become less significant) if we changed
$\Delta v$ to $2000$ km s$^{-1}$ ($P=0.25$) or $4000$ km s$^{-1}$ ($P=0.21$).
These numbers could be misleading, however, 
since even a profound QSO influence
might be difficult to detect with significance in a sample so small.
More convincing is the fact that the QSOs
are too faint to alter the ionization
balance of material that lies $\sim 30$ proper Mpc away (i.e.,
$\Delta v=2000$ km s$^{-1}$; see, e.g., Adelberger~2004).
Their ionizing radiation could explain the galaxies' weak HI absorption
only if the QSOs' redshifts differ significantly from our estimates.

A search for other explanations would have to explore possible
faults with the data themselves.  Precise measurements
of a galaxy's redshift are required to recognize that it
is associated with a narrow absorption feature in a QSOs spectrum.
Any errors in the redshift
will tend to move the galaxy away from the 
feature, artificially reducing the apparent absorption.
If such errors were responsible for galaxies with weak
absorption, one would expect those galaxies always to
have strong absorption features nearby.  
Figure~\ref{fig:check_gpegals} shows the nearby absorption features
for all nine of the galaxies from Figures~\ref{fig:gpegals} 
and~\ref{fig:compare_juna} that had transmissivity
$\bar f_{1{\rm Mpc}}>0.5$.
Strong nearby HI absorption is present for about half of the galaxies,
but systematic wavelength errors could make our data
consistent with the windless simulation only if their magnitude
is much worse than the value $\sigma_v=60$ km~s$^{-1}$ 
that we estimate from repeated observations of individual galaxies.

\section{CHARACTERISTICS OF POSSIBLE SUPERWIND GALAXIES}
\label{sec:characteristics}
The previous sections discussed two characteristics of
intergalactic absorption that might reflect the influence
of a nearby galaxy's superwind:  unusually strong metal
absorption or unusually weak HI absorption.  The two
are contradictory, since metal lines are almost always
produced by gas with a large HI column density.
The association of either with superwinds could be strengthened
if their presence was correlated with properties of the nearby galaxy.

\begin{figure}
\plotone{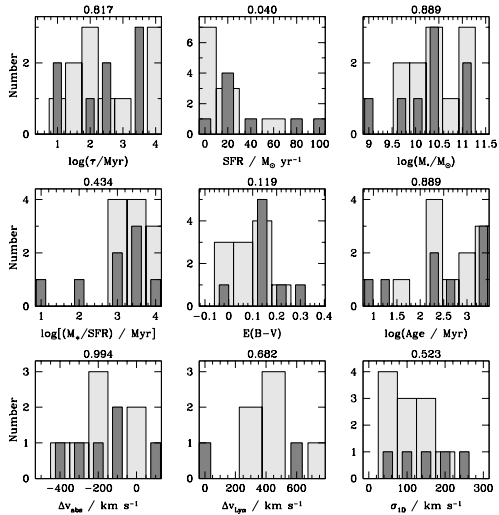}
\caption{
Characteristics of galaxies with little nearby intergalactic HI
($\bar f_{1{\rm Mpc}}>0.5$).
The distributions of these galaxies' properties
are shown with dark bars; the distributions for
the rest of the $<1h^{-1}$ Mpc NIRSPEC
sample are shown in the light histogram.
The number above each panel shows the approximate
Kolmogorov-Smirnov probability of observing the data if
the two distributions were drawn from the same parent population.
This probability is calculated with an approximation 
(subroutine ``probks'' in Press et al. 1992) that
becomes poor for very small sample sizes (as, e.g., in the
bottom center panel.)
The abscissae show:  $\tau$ --- time constant in the assumed
exponential star-formation history ${\rm SFR}\propto e^{-t/\tau}$;
SFR --- star formation rate;
$M_\ast$ --- stellar mass;
$E(B-V)$ --- dust reddening, for a Calzetti (2000)
dust curve; 
Age --- estimated time since beginning of current episode of
star-formation;
$\Delta v_{\rm abs}$ --- velocity difference
between galaxy's near-IR nebular redshift and its interstellar absorption lines;
$\Delta v_{{\rm Ly-}\alpha}$ --- velocity difference between
the near-IR nebular redshift and the Ly-$\alpha$ emission line (which was usually
not detected); $\sigma_{1D}$ --- 
velocity width of the near-IR nebular emission lines.
\label{fig:fluxgpegals}
}
\end{figure}

Figure~\ref{fig:fluxgpegals} compares various properties of
galaxies with $\bar f_{1{\rm Mpc}}<0.5$ (lighter bars)
and $\bar f_{1{\rm Mpc}}>0.5$ (darker bars).
Star-formation histories and stellar masses $M_\ast$
were estimated by fitting model spectral-energy distributions
to the galaxies' observed photometry ($U_nG{\cal R}JK_s$+{\it Spitzer}-IRAC 
3.6, 4.5, 5.8, $8.0\mu$m for galaxies
in Q1700; $U_nG{\cal R}JK_s$ for galaxies in Q1623 and Q2343;
and $U_nG{\cal R}K_s$ for the galaxy in Q2346).
The models assumed exponentially declining star-formation rates
with varying time constants 
and ages, and each was subjected to
a varying amount of reddening by dust that followed a Calzetti (2000) law.
See Shapley et al. (2005) and Erb et al. (2005, in preparation) for details.
Velocity differences between the galaxies' interstellar absorption lines,
Lyman-$\alpha$ emission line, and nebular emission lines were calculated
as in \S~\ref{sec:data}.  The rms widths of the nebular lines
were estimated by fitting the each line's profile with a Gaussian
and subtracting the resolution in quadrature.
Further details can be found in Erb et al. (2004).
The data in these figures are available in Table~\ref{tab:gpegals}.
Numbers above each panel give the approximate significance of
the difference in distributions for the two galaxy populations,
calculated with the routine ``probks'' of Press et al. (1992).
Small numbers correspond to more significant differences.
None of the differences are especially significant given the
number of properties examined, although galaxies with weak nearby 
HI absorption may have somewhat higher star-formation rates.

\begin{figure}
\plotone{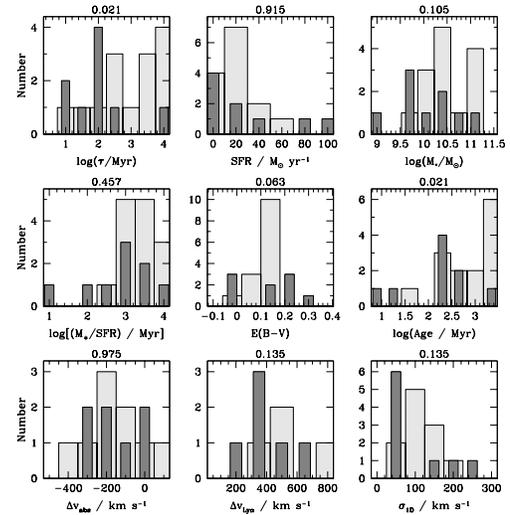}
\caption{
Similar to Figure~\ref{fig:fluxgpegals}, except here we compare
galaxies with and without substantial nearby intergalactic 
CIV absorption.  Dark bars are for galaxies with $N_{\rm CIV}>10^{13}$ cm$^{-2}$,
where $N_{\rm CIV}$ is the total detected intergalactic
CIV column density (cm$^{-2}$) within $200$ km s$^{-1}$ of the galaxy redshift,
and light bars are for galaxies with $N_{\rm CIV}<10^{13}$ cm$^{-2}$.
\label{fig:civgpegals}
}
\end{figure}

Figure~\ref{fig:civgpegals} is similar, except here the sample
is divided by the total detected CIV column density
within $200$ km s$^{-1}$ of the galaxy redshift.
Galaxies with $N_{\rm CIV}>10^{13}$ cm$^{-2}$ are indicated with
darker bars, the remainder with lighter bars.  This differences
in this figure may be marginally significant.  Galaxies near intergalactic CIV systems
appear to have SEDs that require younger ages, shorter time constants,
and a wider range of dust reddening.  It would be a mistake to
interpret this too literally: 
the parameters have large and correlated uncertainties and
our assumed star formation
histories are grossly oversimplified.
The figure suggests, however, that there are genuine differences
between the SEDs of galaxies with and without nearby CIV absorption.
The main empirical difference is a bluer mean optical-to-near-IR color
for galaxies near CIV systems, which our SED fitting interprets
as an excess of $A$ stars relative to $F$ and $G$ stars.

\begin{figure}
\plotone{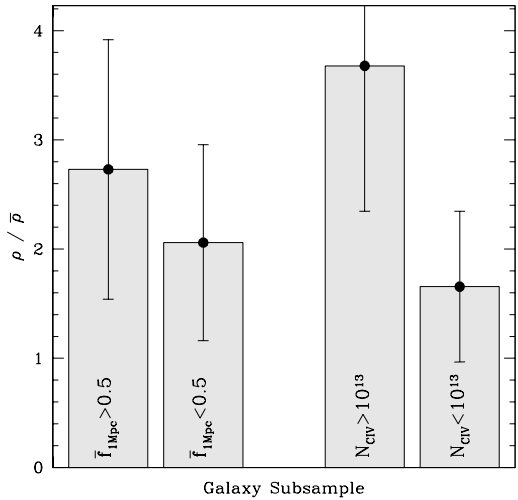}
\caption{
Average large-scale environments for different types of galaxies
in the $<1h^{-1}$ Mpc NIRSPEC sample of figure~\ref{fig:gpegals}.
The environment is quantified through the relative local
galaxy density
$\rho/\bar\rho$, which we calculate by counting the number of
neighboring galaxies in the full sample
within $\Delta\theta=200''$, $\Delta v=500$ km s$^{-1}$
of each galaxy in the subsample
and dividing by the number expected in the absence of clustering.
Galaxies near strong CIV absorbers 
($N_{\rm CIV}>10^{13}$ cm$^{-2}$)
tend to reside in denser environments, possibly indicating
that the CIV originates outside of the galaxy.
In contrast, the prevalence of galaxies with little nearby 
Ly-$\alpha$ absorption does not appear to be strongly influenced
by environment.
\label{fig:compare_densities}
}
\end{figure}

Missing from figures~\ref{fig:fluxgpegals} and~\ref{fig:civgpegals}
is any indication of the large-scale environments that contain
the different galaxy types.  
Shot noise prevents us from estimating the local galaxy density
near any single object, but we can estimate the mean galaxy density
around an ensemble of objects by counting their observed number of
galaxy neighbors and dividing by the number expected in the
absence of clustering.  Figure~\ref{fig:compare_densities} shows
the result for the different galaxy ensembles discussed above.
Galaxies in denser environments are more likely to have 
detectable intergalactic CIV absorption within $1h^{-1}$ comoving Mpc,
but local galaxy density does not seem to affect whether
a galaxy has  $\bar f_{1{\rm Mpc}}>0.5$ or $\bar f_{1{\rm Mpc}}<0.5$.

\section{DISCUSSION}
\label{sec:discussion}
\subsection{Superwinds at $2\simlt z\simlt 3$}
Galaxies would presumably be associated with QSO absorption lines even if there
were no winds.  This subsection discusses how strongly (or weakly) our observations
support the existence of superwinds around the galaxies.  We consider below the possibility
that intergalactic metals were produced at very high redshifts ($z\sim 10$).

The blueshifted absorption and redshifted Lyman-$\alpha$ emission in high-redshift
galaxies' spectra is the strongest evidence for winds from the galaxies.
It is expected in a wide range of models in which the
stars are surrounded by outflowing gas (e.g., Tenorio-Tagle et al. 1999;
Zheng \& Miralda-Escud\'e 2002) and is not (as far as we know)
expected in any other class of realistic models.
The statistical significance is high, since the same pattern has been observed
in the spectra of dozens of galaxies (e.g., figure~\ref{fig:check_dv}).
This observation only shows that the outflowing gas lies somewhere outside
the stellar radius, however.  It does not imply that the outflow will be driven
into the surrounding IGM.

The strength of CIV absorption at $b=40$ kpc (Figure~\ref{fig:specsum}) may suggest
that the outflows have normally advanced to this radius,
but even so this is only about half way to the virial radius
(for $M_{\rm total}\sim 10^{12}M_\odot$ in the Lambda-CDM cosmology
favored by WMAP; Spergel et al. 2003).
The minimum allowed velocity range of the gas at $b=40$ kpc
($\sim 300$ km s$^{-1}$) is not large enough to guarantee escape.
In any case the similarity of the CIV absorption strength in Lyman-break Galaxies' (LBG)
spectra and at impact parameter $b=40$ kpc could just be a coincidence; it does
not
rule out the idea that the gas at 40 kpc is falling into or
orbiting within the galaxy's potential.
Although anisotropies in the galaxy/CIV-system correlation function seem to disfavor
infall and favor rapid outflows for the origin of the CIV,
CIV systems tend to be found near galaxies which may be too young
to have driven winds far into their surroundings.
The small number of distinct galaxy-CIV pairs leaves the situation unclear.

Measuring the spatial correlation of galaxies and metals on large (Mpc) scales
would seem to provide a powerful way to distinguish between different
scenarios for intergalactic metal enrichment.  If the metals
were produced in LBGs, they should have a spatial bias\footnote{on scales larger than
the maximum wind radius} similar to the
galaxies' bias of $b\sim 2.5$ at $z=3$ (Adelberger et al. 2005).
They would have $b=1$ if they were produced at any redshift by
the numerous dwarf galaxies that are $1\sigma$
fluctuations, and $b\sim 1.9$ at $z=3$ if (as envisioned by Madau, Ferrara, \& Rees 2001)
they were produced at $z=9$ by galaxies that were $2\sigma$ fluctuations.
These estimates of the bias exploit Mo \& White's (1996)
high-redshift (i.e., $\Omega\sim 1$) approximation
\begin{equation}
b(z) \simeq 1 + \frac{(\nu^2-1)(1+z)}{1.69(1+z_i)}
\label{eq:mw}
\end{equation}
for the bias at redshift $z$ of objects that were $\nu$-sigma fluctuations
at the earlier redshift $z_i$.
The galaxy-metal cross-correlation length would therefore be equal to the galaxy-galaxy
correlation length if the metals were produced by LBGs and smaller
for the other two cases.  Unfortunately the implied bias for enrichment at $z\sim 10$
depends sensitively on the unknown value of $\nu$.  The parameter $\nu$ would not
have to be very different from the default assumption of Madau, Ferrara, \& Rees (2001)
to make the bias exactly equal to the bias for enrichment at $z\sim 3$.  
In any case, we have measured the galaxy-CIV
correlation length, not the galaxy-metal correlation length.
Our measured cross-correlation function therefore does 
not have an unambiguous interpretation.

There are only 9 galaxies in our sample with precise (near-IR nebular line)
redshifts and little HI within $1h^{-1}$ comoving Mpc, and their statistical
significance is even lower than one might imagine.  There are two reasons.
First, as discussed above, the measurement is difficult.  To determine
whether a galaxy lies within $\simlt 1h^{-1}$ Mpc of a narrow absorption line
in a QSO spectrum, one needs
a good understanding of many possible sources of random and systematic error
in the estimated redshifts.
In about half the cases there is no absorption line anywhere near the galaxy redshift,
but in the remainder the galaxies could conceivably be associated with nearby
absorption lines
if redshift errors were somewhat larger than we estimate.
Reducing the number of galaxies with little nearby HI by 50\% would
begin to make our observations consistent with the predictions of windless SPH simulations.
Second, 5 of the 9 galaxies lie within a single field, Q1623, which contains one of
the largest known concentrations of QSOs with $2\simlt z\simlt 3$.
This is a consequence of the dense spectroscopic sampling we obtained in the field,
but it raises the possibility that our result might have been different
had we surveyed a more representative part of the universe.
The spatial clustering strength of the high-redshift galaxies
in this field is certainly not typical, for example (Adelberger et al. 2005b).
                                                                                
Even without its questionable statistical significance, the apparent lack
of HI near some galaxies would not have a straightforward interpretation.
This is due to the complexity of the interaction
between winds and galaxies' inhomogeneous surroundings.
Analytic models treat this in an extremely crude way, and
existing numerical simulations are unable to resolve either
the shock fronts or the instabilities that result when the hot wind
flows past cooler intergalactic material.  As a result it is unclear
if winds would destroy the HI near galaxies.
Other effects that simulations do not resolve (e.g., cooling
instabilities) might reduce the covering fraction of HI near galaxies.
Even if there were low-density regions near galaxies in SPH simulations,
they might not be recognized since the density is generally estimated
by smoothing over the few dozen nearest particles, particles which
(by definition) are likely to lie outside of any local underdensity.
Finally, photoionization could play some role.
Kollmeier et al. (2003) and Adelberger et al. (2003)
have shown that the galaxies' radiation
cannot destroy all the HI within $1h^{-1}$ comoving Mpc
without overproducing the UV background, but photoionizing the
weaker clouds around one third of the galaxies might be possible.
It therefore remains unclear if the lack of HI near some galaxies is
a unique or even an expected signature of superwinds.
The weak correlations between galaxy properties and
the strength of nearby intergalactic HI absorption
may suggest that the
HI absorption is unrelated to winds.

\subsection{Metal enrichment at $z\sim 10$}
\begin{figure}
\plotone{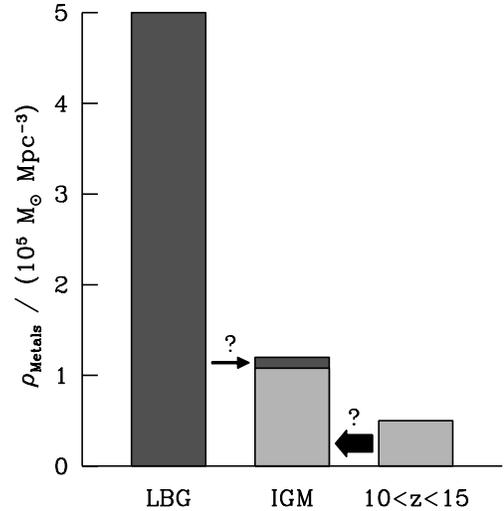}
\caption{
The amount of metals produced in massive galaxies at $z\sim 2$ (LBG) compared to
the amount of metals in the IGM at $z\sim 2$ and to the estimated maximum amount of
metals produced at $10<z<15$.  %The amount of metals produced by massive galaxies
%at $z\sim 2$ was calculated by assuming that each $100 M_\odot$ of star formation
%produced $1 M_\odot$ of metals and scaling from the $z\sim 2$ stellar-mass
%density measured by Dickinson et al. (2003) and Rudnick et al. 2003.  An upper limit
%to the amount of metals produced at $10<z<15$ was derived by assuming a constant
%comoving star-formation density at all redshifts $2<z<15$.  This implies
%that $\sim 10$\% of the stars that exist at $z=2$ were formed
%at $10<z<15$.  It is an upper limit because the actual star-formation density
%at $z\gg 2$ appears to be significantly lower than the star-formation density
%at $z\sim 2$.  The amount of metals in the IGM was calculated by multiplying
%the estimated intergalactic carbon density at $z\sim 2$, $\Omega_C\sim 2\times 10^{-7}$
%(Schaye et al. 2003), by the critical density.
Star formation at $10<z<15$ probably does not produce enough metals to account
for the observed enrichment at $z\sim 2$.  Even if it did,
the fraction of metals that escape into the IGM would have to be
$\sim 100$ times higher at $10<z<15$ than at $z\sim$2--3
for 90\% of the intergalactic metals at $z\sim 2$ to have been
produced at $10<z<15$.
\label{fig:metals}
}
\end{figure}

Although our observations do not provide unequivocal evidence for 
superwinds at $2\simlt z\simlt 3$, it would be a mistake to
conclude that they favor intergalactic metal enrichment at very high redshifts.
In the few cases where the two scenarios have clearly different
predictions (e.g., anisotropies in the galaxy-CIV correlation function
on small scales), the observations seem to favor lower-redshift enrichment.
In addition, enrichment at very high redshifts seems
implausible on other grounds.
Figure~\ref{fig:metals} illustrates a major problem:
stars that form before $z\sim 10$ probably produce an inconsequential
amount of metals compared to the stars that form afterward,
so their metals could be dominant in the IGM only if 
virtually no metals were able to escape into the IGM at later times.
In order for 90\% of the
intergalactic metals observed at $z\sim 3$ to have
been produced at $10\simlt z\simlt 15$, for example,
supernovae ejecta would have to be at least 100 times
more likely to escape their galaxy at $z=10$
than at $z=3$.  This follows from the fact that the
total intergalactic metal density at $z\sim 2$, 
$\rho_{\rm met}\simeq \rho_{\rm cr}\Omega_{\rm met}\simeq 1.2\times 10^5 M_\odot {\rm Mpc}^{-3}$
for $\Omega_{\rm met}\simeq 4.4\Omega_C$ and
$\Omega_{C}\simeq 2.3\times 10^{-7}$ (Schaye et al. 2003),
would be roughly twice the total metal production at $10<z<15$
and five times smaller than the metal production
at $2\simlt z\simlt 5$ if the cosmic star-formation density were
roughly constant for $2<z<15$,  the stellar density at
$z\sim 2$ were $\sim 5\times 10^7 M_\odot {\rm Mpc}^{-3}$ (e.g.,
Dickinson et al. 2003; Rudnick et al. 2003), and
$100M_\odot$ of star formation produced $1M_\odot$ of metals.

%\clearpage

Is the metal escape fraction likely to be 100 times larger at
$z\sim 10$ than $z\sim 3$?  Two factors are commonly thought to increase
the likelihood of escape at high redshift:  galaxies are less massive and
the intergalactic medium lies closer to them than at later times when
the universe has expanded more.
Both are offset by competing effects, however, and it is not
clear to us that very high redshift galaxies have much of an advantage at all.

First consider the galaxy masses.  The gravitational potential energy of
a galaxy scales as $M^2/r$, or $M^{5/3}$, while the energy released
by a galaxy's supernovae scales as $f M$, where $f$ is the
efficiency of star formation.  If $f$ were independent of $M$,
smaller galaxies would be more easily unbound by supernova explosions.
This is the standard assumption, 
but in fact $f$ increases with $M$ and the situation is not clear cut.
For example, compare the star-formation efficiencies of Lyman-break
galaxies at $z\sim 3$ to those of galaxies at $z\sim 10$.
The ratio of stellar to baryonic mass for Lyman-break galaxies at $z\sim 3$ 
is $M_\ast/M_b\sim 0.1$ (Adelberger et al. 2004), which implies $f\sim 0.1$.
An upper limit to the star-formation efficiency at $z=10$ can
be derived by assuming that
the star-formation density is constant for $2<z<15$.  (In fact it
appears to be significantly lower at $z\simgt 5$; Bunker et al. 2004.)  
In this case
the stellar density
at $z=10$ will be $\Omega_\ast/\Omega_b=0.001$, scaling from the $z\sim 2$ stellar density
estimated by Dickinson et al. (2003) and Rudnick et al. (2003), while the
fraction of baryons in the halos with $v_c\simgt 10$ km s$^{-1}$ that are
able to host star formation (Dijkstra et al. 2004) is $\Omega_h/\Omega_b \sim 0.08$.  
That implies an upper limit to the star-formation
efficiency of $f\simeq\Omega_\ast/\Omega_h\sim 0.01$ 
for typical galaxies at $z\sim 10$.
The order-of-magnitude increase in star-formation efficiency at lower redshift
removes any advantage higher-redshift galaxies might gain from their smaller masses.
A similar argument shows that the star-formation efficiencies of LBGs
are far higher than those of smaller galaxies at similar redshifts $2\simlt z\simlt 3$.

\begin{figure}
\plotone{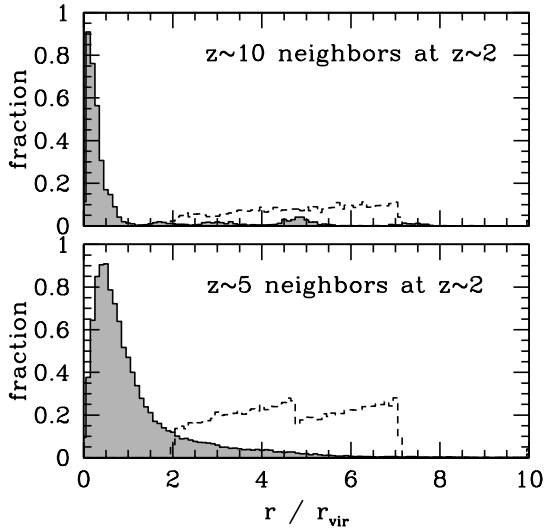}
\caption{
Distance to the nearest massive ($M\simgt 10^{11}M_\odot$)
halo at $z\sim 2$ for all GIF-LCDM simulation
particles whose distance $r$ to the nearest halo satisfied
$2r_{\rm vir} < r < 1h^{-1} {\rm comoving Mpc}$ at $z=10$ (top) or $z=5$
(bottom).  Dashed lines show the particles' original
(higher redshift) distribution
of $r/r_{\rm vir}$; the solid shaded histograms show the
distribution at $z\sim 2$.  (The nearest halo and its virial
radius are usually significantly different at the initial and final times.)
These particles initially lie
at larger radii than those expected for the metals ejected
by very high redshift winds, yet they mostly end up inside halos
by $z\sim 2$.  This suggests that the metals ejected at $z\sim 5$ and $z\sim 10$
will generally also lie in massive halos at $z\sim 2$, not in the IGM.
\label{fig:haloneighbors}
}
\end{figure}

Now consider the supposed benefit from the small size of the universe at $z\sim 10$.
If supernova ejecta at $z\sim 10$ extended from their galaxies
into the receding Hubble flow, their filling fraction
at lower redshift would be boosted by the subsequent expansion of the universe.
In fact, however, the ejecta will not normally reach the Hubble flow.  Instead they
are likely to be swept back into their galaxy by the ongoing process of structure formation.
To illustrate the point, we show in Figure~\ref{fig:haloneighbors}
the distance to the nearest halo of mass $M\simgt 10^{11} M_\odot$
at $z=2.12$ for every particle
in the GIF-LCDM simulation (Kauffmann et al. 1999)
that lay within $1h^{-1}$ comoving Mpc of a galaxy (i.e., halo)
at $z=10$ or $z=5$.  The result
is stark:  the overwhelming majority of these particles end up inside
the virial radius of a galaxy at $z\sim 2$.  Since the stalled
ejecta of galaxies at $z=10$ or $z=5$ will be swept along by the
movements of the material that surrounds the galaxies, they should
largely end up inside galaxies at $z\sim 2$ as well.
This exercise may be slightly misleading,
since the GIF-LCDM simulation only resolves halos of mass $M\simgt 10^{11}M_\odot$
(i.e., $\sim 4\sigma$ fluctuations at $z\sim 10$),
not the smaller halos believed to be most responsible for polluting the
IGM at $z\sim 10$.  However, a simple calculation shows that a significant fraction
of the metals from the lower-mass progenitors should also end up inside galaxies
at $z\sim 2$.  The estimated stalling radius for winds from small galaxies
at $z\sim 10$ is $\sim 100$ comoving kpc (e.g., Madau et al. 2001), which
is the Lagrangian radius for a halo of mass $1.7\times 10^8 M_\odot$.  If these galaxies'
typical descendants at $z\sim 2$ have masses significantly larger than this,
the metals will likely have been swept inside them.  According to the extended
Press-Schechter formalism,
$\sim 85$\% of the galaxies at $z=2$
that descended from $2\sigma$ fluctuations at $z=10$ will have masses that exceed
this threshold by an order of magnitude
(see, e.g., equation 2.16 of Lacey \& Cole 1993).  A substantial fraction of the metals
produced at $10\simlt z\simlt 15$ should therefore be locked inside galaxies by $z\sim 2$.
Once there, some of them are likely to cool further, fall towards the center,
and disappear from view.
Since the total
metal production at $10\simlt z\simlt 15$ is at best comparable
to the observed metal content of the IGM at $z\sim 2$ (Figure~\ref{fig:metals}),
this suggests that the intergalactic metallicity at $z\sim 2$ must receive a significant
contribution from some other source.
A corollary is that the metal content of the IGM would drain into galaxies and
{\it decrease} over time if it were not continually replenished.  The observed
constancy of the IGM metallicity (e.g., Schaye et al. 2003) therefore also seems to require
metals to escape from galaxies at lower redshifts.\footnote{If one
removes the assumption that the metals injected into the IGM 
at a given redshift must lie near galaxies that existed at that
redshift, then it is possible to find a spatial distribution of intergalactic
metals that matches intergalactic absorption-line statistics
and does not drain into galaxies at a significant rate
for $2\simlt z\simlt 4$ (e.g., Schaye et al. 2003).
The removal of this assumption seems dubious to us, however.
Metals could exist in parts of the IGM that were never near galaxies
only if they were produced by Population III stars, presumably,
and the amount of Population III star formation would have to be
inordinately large to match the observed intergalactic metallicity
at $z \sim 2$ (e.g., Figure~\ref{fig:metals}; see also
Aguirre et al. 2004).}

\section{SUMMARY}
\label{sec:summary}
We reviewed the status of our search for direct
evidence of large-scale outflows around UV-selected
star-forming galaxies at $2\simlt z\simlt 3$.
These are our principal observations:

$\bullet$ The gas that lies within 40kpc of LBGs produces extremely strong absorption
lines ($N_{\rm CIV}\gg 10^{14}$ cm$^{-2}$) in the spectra of background galaxies and QSOs
(Figure~\ref{fig:specsum}).
The large equivalent widths of the CIV absorption in low-resolution spectra imply that the
absorbing material has range of velocities of at least $\Delta v=260$ km s$^{-1}$.
The absorption produced by this gas is similar to the interstellar absorption seen
in LBGs' spectra, suggesting the LBGs' outflowing ``interstellar'' gas may actually
lie at radii approaching 40kpc.
                                                                                                          
$\bullet$ For roughly half of the LBGs,
CIV absorption lines with column density $N_{\rm CIV}\sim 10^{14}$ cm$^{-2}$
are observed out to impact parameters of $\sim 80$ kpc
(Figures~\ref{fig:specsum} and~\ref{fig:civhist}).
This implies that roughly one-third of all ``intergalactic'' absorption lines
with $N_{\rm CIV}\simgt 10^{14}$ cm$^{-2}$ are produced by gas that lies within
$\sim 80$ kpc of an LBG.  Galaxies too faint to satisfy our selection criteria
could easily account for the remainder.  In some cases the absorbing gas
has substructure on half-kpc scales (Figure~\ref{fig:smallscalestructure}).
                                                                                                          
$\bullet$  The cross-correlation function of galaxies and CIV systems with
$N_{\rm CIV}\simgt 10^{12.5}$ cm$^{-2}$
appears to be the same as the correlation function of galaxies
(Figure~\ref{fig:r0gc_vs_nciv}).
This implies that CIV systems and galaxies reside in similar parts of
the universe and is consistent with the idea that they are largely
the same objects.  We also find a strong association of OVI systems with
galaxies (Figure~\ref{fig:q1700metals}).
On small scales the
redshift-space cross-correlation function is highly anisotropic
(Figure~\ref{fig:civ_nirspec_xigc}).
The metal-enriched gas must therefore have large velocities
relative to nearby galaxies.  The required velocities appear to
exceed the galaxies' velocity dispersions
but are similar to the galaxies' observed outflow speeds.
                                                                                                          
$\bullet$ In contradiction to the earlier result of Adelberger et al. (2003),
we find that the gas within $1h^{-1}$ comoving Mpc of LBGs
usually produces strong Ly-$\alpha$ absorption in the spectra
of background galaxies (Figure~\ref{fig:fbar_vs_r}).  The absorption is weak (mean transmitted flux
within $1h^{-1}$ Mpc of $\bar f_{1{\rm Mpc}}>0.5$)
in only about one case out of three (Figure~\ref{fig:compare_juna}).
Even so, the weakness of the absorption
in these cases remains difficult to understand.  Since high-redshift galaxies reside in
dense parts of the universe, with large amounts of hydrogen and short
recombination times, one would expect them to be surrounded by
large amounts of HI.
The HI has presumably collapsed into clouds and sightlines
to the background QSOs might occasionally miss every cloud near a galaxy,
but the SPH simulations of Kollmeier et al. (in preparation) suggest that
the chance of this is very low (Figure~\ref{fig:compare_juna}).

$\bullet$ We were unable to identify any statistically significant differences
in age, dust reddening, stellar mass, kinematics, or environment
between galaxies with weak nearby HI absorption and the rest, although
galaxies with weak absorption may have higher star-formation rates
(Figure~\ref{fig:fluxgpegals}).
Galaxies near intergalactic CIV systems appear to reside in relatively dense
environments (Figure~\ref{fig:compare_densities}) and to
have distinctive SEDs characterized by blue colors and young ages
(Figure~\ref{fig:civgpegals}).

As discussed in \S~\ref{sec:discussion}, none of these observations
provide unequivocal evidence for superwinds at $z\sim 2$--3.  
This is mostly because it is unclear
how superwinds would affect the correlation between 
galaxies and intergalactic absorption lines.
Metals and galaxies would lie near each other
at $z\sim 3$ even if there were no active outflows, for example.
We argued in \S~\ref{sec:discussion} that star-formation at
$10<z<15$ may not produce enough metals to account for
the intergalactic metallicity at $z\sim 2$--3, and that many of
these metals would be buried inside galaxies by $z\sim 2$--3,
but our simple arguments need to be checked with numerical simulations.
The interpretation of our data will remain ambiguous until that time.
The arguments presented in \S~\ref{sec:discussion} should make it
clear that metal enrichment at $5\simlt z\simlt 15$ will be much
harder to rule out than metal enrichment at $10\simlt z\simlt 15$.

There is only one case where the data
seem easy to interpret and we are limited by the small size of
our sample.  
This is the redshift-space distribution of metals within $\sim 200$ comoving kpc
of galaxies.  Mapping the velocity offsets between galaxies' stars
and their metals as a function of impact parameter would help
show if the detected metals are flowing out of, falling into, or orbiting
within the galaxies' halos.  The velocity offsets can be mapped by obtaining
higher resolution spectra of more close pairs (e.g., the right
panel of figure~\ref{fig:schematic}) or by obtaining more near-IR nebular redshifts
for galaxies near the QSO sightline (e.g., Figure~\ref{fig:civ_nirspec_xigc}).
That seems the sensible way to proceed.

\bigskip
\bigskip
Trenchant criticism from P. Madau, P. Petitjean, M. Rauch, J. Schaye,
and an anonymous referee
improved this paper significantly.  We benefited as well from
conversations with T. Abel, J. Kollmeier, J. Miralda-Escud\'e, 
D. Weinberg and S. White.
W. Sargent and M. Rauch gave us generous access to their QSO spectra.
J. Kollmeier, D. Weinberg, N. Katz, and R. Dav\'e allowed
us to show some of their unpublished results.
KLA and AES were supported by
fellowships from the Carnegie Institute of Washington and the Miller Foundation.
DKE, NAR, and CCS were supported
by grant AST 03-07263 from the National Science Foundation
and by a grant from the Packard Foundation.
During the course of our observations
we frequently consulted the NASA/IPAC Extragalactic Database (NED),
which is operated by the Jet Propulsion Laboratory, California Institute
of Technology, under contract with the National Aeronautics and Space
Administration.
We are grateful that the people of Hawaii allow astronomers
to build and operate telescopes on the summit of Mauna Kea.

\end{document}